\documentclass[referee,a4paper,12pt,traditabstract]{swsc} 

\usepackage{graphicx}
\usepackage{txfonts}
\usepackage{subfigure}
\usepackage{epstopdf}
\usepackage[displaymath,mathlines]{lineno}
\usepackage[authoryear,round]{natbib}
\usepackage[backref]{hyperref}
\usepackage{url}

\hypersetup{colorlinks=true,citecolor=cyan,urlcolor=cyan,linkcolor=blue}

\divide\pdfpxdimen by 96
\begin{document}


   \title{The GOES-R EUVS Model for EUV Irradiance Variability}

   \titlerunning{The EUVS Model}

   \authorrunning{Thiemann et al.}

   \author{G. Wuchterl
          \inst{1}
          \and
          C. Ptolemy\inst{2}\fnmsep\thanks{Just to show the usage
          of the elements in the author field}
          }

\author{E. M. B. Thiemann\inst{1}
	\and 
		F. G. Eparvier\inst{1}
		\and 
		D. Woodraska\inst{1}
		\and 
		P. C. Chamberlin\inst{1}
		\and
		J. Machol\inst{2,3}
		\and 
		T. Eden\inst{1}
		\and 
		A. R. Jones\inst{1}
		\and 
		R. Meisner\inst{1}
		\and 
		S. Mueller\inst{1}
		\and
		M. Snow\inst{1}
		\and 
		R. Viereck\inst{4}
		\and
		T. N. Woods\inst{1} 
}

\institute{ Laboratory for Atmospheric and Space Physics, University of Colorado, Boulder, CO, USA
	\and
	Cooperative Institute for Research in Environmental Sciences, University of Colorado, Boulder, CO, USA
	\and
	National Centers for Environmental Information, National Oceanic and Atmospheric Administration, Boulder, CO, USA
	\and
	Space Weather Prediction Center, National Oceanic and Atmospheric Administration, Boulder, CO, USA}


 
  \abstract
   {The Geostationary Operational Environmental Satellite R (GOES-R) series of four satellites are the next generation NOAA GOES satellites. Once on orbit and commissioned, they are renamed GOES 16-19, making critical terrestrial and space weather measurements through 2035.  GOES 16 and 17 are currently on orbit, having been launched in 2016 and 2018, respectively.  The GOES-R satellites include the EUV and X-ray Irradiance Sensors (EXIS) instrument suite, which measures calibrated solar irradiance in 8 lines or bands between 25 and 285 nm with the Extreme Ultraviolet Sensors (EUVS) instrument.  EXIS also includes the X-Ray Sensor (XRS) instrument, which measures solar soft X-ray irradiance at the legacy GOES bands.  The EUVS measurements are used as inputs to the EUVS Model, a solar spectral irradiance model for space weather operations that predicts irradiance in twenty-two 5 nm wide intervals from 5 nm to 115 nm, and one 10 nm wide interval from 117 to 127 nm at 30 second cadence.  Once fully operational, NOAA will distribute the EUVS Model irradiance with 1 minute latency as a primary space weather data product, ushering in a new era of rapid dissemination and measurement continuity of EUV irradiance spectra.  This paper describes the EUVS Model algorithms, data sources, calibration methods and associated uncertainties.  Typical model (relative) uncertainties are less than $\sim$5\% for variability at time-scales longer than 6 hours, and are $\sim$25\% for solar flare induced variability.  The absolute uncertainties, originating from the instruments used to calibrate the EUVS Model, are $\sim$10\%.  Examples of model results are presented at both sub-daily and multi-year timescales to demonstrate the model's capabilities and limitations.  Example solar flare irradiances are also modeled.  
}        


   \maketitle

\section{Introduction} \label{sec_intro}

	Solar extreme ultraviolet (EUV, 10-121 nm) irradiance is the primary energy input into the Earth's upper atmosphere at low to mid latitudes and at all latitudes during geomagnetically quiet periods.  By ionizing gases, EUV radiation creates the ionosphere and heats the thermosphere.  The solar EUV irradiance varies significantly due to the evolving 11-year solar cycle, 27-day solar rotation period and transient solar flares; resulting in corresponding changes in the density, temperature and composition of the thermosphere and ionosphere.  Additionally, both the solar spectrum and atmospheric gas absorption cross-sections are highly structured at EUV wavelengths, influencing the altitudes at which solar EUV radiation is absorbed.    EUV-induced changes in the thermosphere can extend to Low Earth Orbit (160-2000 km), directly impacting satellite drag, where a hotter, denser thermosphere exerts a greater drag force on the satellites orbiting within it \citep{jacchia1959two}. With regard to the ionosphere, EUV-induced changes modulate its index of refraction, impacting transionospheric communication and navigation signals \citep{davies1990ionospheric}. The response of both the ionosphere and the thermosphere to EUV changes are relatively fast, with the ionosphere responding near instantaneously and the thermosphere responding in 2-4 hours \citep[e.g.][]{mendillo1974behavior,qian2010flare}. As such, near real time solar EUV spectral irradiance information is needed for accurate space weather forecasting. To meet this demand, the National Oceanic and Atmospheric Administration (NOAA) has included an operational (continuous, high time cadence, low latency) EUV irradiance data product as part of its Geostationary Operational Environmental Satellite R (GOES-R) series program.  The GOES-R satellites are scheduled to make observations from 2016-2035, providing nearly two decades of real-time continuous solar EUV irradiance measurements, and changing the paradigm for the availability and dissemination of spectral EUV irradiance data.
	
	The NOAA GOES-R series program is the latest iteration of the GOES satellite constellation, and consists of four satellites, each carrying a suite of identical instruments designed to monitor terrestrial and space weather continuously.  Once commissioned, the GOES-R series satellites will become GOES 16-19, with the first in the series, GOES-16, having launched on 19 November 2016, and the last in the series, GOES-19, anticipated to launch in the early 2020s and to be operational until at least 2035.  The GOES satellites have monitored solar soft X-ray irradiance in two bands since their inception in 1975 with the X-Ray Sensor (XRS) instruments, and GOES 13-15 have measured EUV irradiance in several bands with the EUV Sensor (EUVS) instruments \citep{viereck2007solar}.  For the GOES-R series satellites, new versions of the XRS and EUVS instruments have been built as part of the EUV and X-ray Irradiance Sensors (EXIS) instrument suite. The EXIS instrument suites were built at the Laboratory for Atmospheric and Space Physics (LASP) at the University of Colorado at Boulder.  Each EXIS suite consists of the EUVS \citep{eparvier2009extreme}, which measures solar emissions at 8 lines or bands between 25 and 285 nm, and the XRS \citep{chamberlin2009next}, which measures soft X-ray irradiance in the legacy 0.1-0.8 nm and 0.05-0.4 nm bands.

 The solar atmosphere is comprised of 4 distinct regions, which are (from nearest to furthest from the surface): the photosphere, chromosphere, transition region, and corona. They can be equivalently categorized according to temperature with the photosphere being the coolest ($\sim$5700 K) and the corona being the hottest ($>$1 MK). Solar radiation emitted from the different regions of the solar atmosphere tends to vary differently as a result of the different processes driving the dynamics within them. For example, bright plasma tends to be concentrated regionally in active region magnetic loops in the corona, whereas in the chromosphere, bright plasma is distributed more uniformly across a magnetic network \citep{antia2003lectures}. 
  
The emission lines observed by EUVS were selected to span a broad range of emission formation temperatures in the solar atmosphere in order to capture a broad range of irradiance variability.  Specifically, EUVS-A measures the He II 25.6 nm, Fe XV 28.4 nm  and the He II 30.4 nm lines; EUVS-B measures the C III 117.5 nm, H I 121.6 nm (Lyman-$\alpha$), C II 133.5 nm and the Si IV/ O IV (blended) 140.5 nm lines; and EUVS-C measures the Mg II emission line core-to-wing ratio \citep[hereafter, the Mg II Index]{heath1986mg} around 280 nm.  These 8 fully-calibrated EUVS measurements (hereafter, the EUVS Measurements) are used as inputs to the EUVS Level 1B solar irradiance model (hereafter, the EUVS Model), which predicts solar EUV spectral irradiance continuously at 30 second cadence and 1 minute latency in twenty-two 5 nm wide intervals from 5-115 nm and a single 10 nm interval from 117-127 nm.

Solar EUV and Far Ultraviolet (FUV, 122-200 nm) irradiance is completely absorbed in the Earth's upper atmosphere, requiring it to be measured by sophisticated space-based instrumentation that is prone to degradation.  This measurement difficulty has resulted in extended periods of time when direct spectral irradiance measurements are unavailable, necessitating models of solar EUV and FUV irradiance to bridge the observational gaps.  Additionally, because calibrations often differ from instrument to instrument, models of solar irradiance provide the capability to estimate irradiance over a long timespan, while avoiding discontinuities that occur at the edges of concatenated datasets due to calibration differences.  

 \citet{hinteregger1981representations} proposed the first widely adopted EUV spectral irradiance variability model, which was based on the Atmospheric Explorer-E EUV Spectrophotometer (AE-E/EUVS) measurements. In his paper, Hinteregger reported two models. The first model used the Fe XVI 33.5 nm and H Lyman-$\beta$ 102.6 nm   emissions to model coronal and chromospheric variability, respectively, with a set of regression coefficients that related these two emissions to the EUV irradiance spectrum. Fe XVI 33.5 nm  and H Lyman-$\beta$ 102.6 nm were only available from AE-E/EUVS during this epoch and therefore this model was of limited use, but the success of the method laid the foundation for models that followed, including the GOES EUVS Model described here.  The second model proposed by  \citet{hinteregger1981representations} used regression coefficients relating the AE-E/EUVS measurements and the 10.7 cm solar radio flux (F10.7) daily and 81-day mean values to estimate the spectrum. Later, \citet{torr1985ionization} re-partitioned the  \citet{hinteregger1981representations}  model spectra into 37 spectral intervals based on the absorption cross-sections of major constituent gases in the Earth's atmosphere.This repartitioning used twenty 5 nm intervals from 5 to 105 nm and then seventeen narrow intervals to model the irradiance of individual emissions lines near regions where atmospheric absorption cross-sections are highly structured. A significant advance occurred when \citet{richards1994euvac} introduced the EUV flux model for Aeronomic Calculations (EUVAC), which  was essentially a re-calibration of the  \citet{torr1985ionization} model coefficients to be consistent with more accurate rocket measurements. Later work by \citet{tobiska1998euv97} and \citet{tobiska2000solar2000} added space-borne model inputs in addition to F10.7; and incorporated new suborbital rocket measurements and data from the San Marco 5 satellite into the model calibration dataset. Regular sub-daily EUV measurements made by the Solar EUV Experiment (SEE) on board the Thermosphere Ionosphere Mesosphere Energetics and Dynamics (TIMED) satellite were incorporated into the Flare Irradiance Spectral Model (FISM) developed by Chamberlin et al. (2007).  \citet{Chamberlin2008flare} incorporated the capability for estimating solar flare irradiance based on the GOES/XRS irradiance into FISM.  FISM was recently updated by \citet{thiemann2017maven} to use measurements from the EUV Monitor (EUVM) on board the Mars Atmosphere and EvolutioN (MAVEN) probe as inputs and included spectral irradiance data from the EUV Variability Experiment (EVE) on board the Solar Dynamics Observatory (SDO) in the model calibration dataset. The Solar Spectral Proxy Irradiance from GOES \citep[SSPRING,][]{suess2016solar} model uses irradiance measurements from GOES-15 as inputs. The EUVS Model presented here is an iteration of these preceding empirical models for solar spectral irradiance, and the first to be implemented in an operational environment.
 
   The long term dataset of EUV spectral irradiance observed by TIMED/SEE provided an opportunity to rigorously test the idea that solar EUV variability can be decomposed into groups of emissions which vary similarly depending on the layer of the solar atmosphere from which they originate. A number of studies investigated these assumptions statistically \citep{kretzschmar2006retrieving, Chamberlin2007flare, de2008solar, amblard2008euv, lilensten2007recommendation, dudok2009finding, cessateur2011monitoring}. Notably, \citet{amblard2008euv} used statistical methods to deconstruct the solar spectrum into elementary components and found that a minimum of 3  elementary spectra are needed to reconstruct the EUV spectral irradiance, representative of the inactive Sun, cool chromosphere and hot corona.  \citet{Chamberlin2007flare} used the long-term TIMED/SEE dataset to determine model error for the integrated 0.1-193 nm band is reduced by $\sim$30$\%$ when space-borne model-inputs are used in lieu of F10.7. Later, \citet{cessateur2011monitoring} used statistical methods to analyze the TIMED/SEE dataset and determined that empirical model error could be reduced by a factor of two using four spectral pass-bands measured from space rather than the commonly used F10.7 and Mg II indices. 

This paper presents the EUVS Model, its coefficients and the methods used to derive them, as well as the model uncertainty.  In Section \ref{sec_model}, the model is described and the model equations are presented.  Section \ref{sec_data} presents the historical data used to compute the model coefficients, the cross-calibration between historical and EXIS measurements, and the associated systematic uncertainty.  Sections \ref{sec_long} and \ref{sec_short} present the long- and short-term model components and their uncertainty; and Section \ref{sec_disc} discusses the overall results.  

The primary purpose of this paper is to show how the EUVS Model coefficients and uncertainties are derived, rather than serve as a reference for their exact values. The coefficients and uncertainties presented here are for GOES-16 at the time of this writing and may change with updates to the GOES-16/EXIS calibration.  Additionally, the coefficients and uncertainties for the GOES 17-19 EUVS Models will possibly differ from those presented here.  As such, current values for the EUVS Model coefficients and uncertainties will be made available via the web address listed in the Acknowledgements section.  

\section{Model Description} \label{sec_model}

The EUVS Model estimates irradiance in the $n^{th}$  wavelength interval, $E_n(t)$, by solving the equation,
\begin{equation}
E_n(t)=E_{n,0}+\sum_{i=1}^{8}j_{i,n}P_i(t) + \sum_{i=1}^{8}k_{i,n}Q_i(t) , \label{eqn_model}
\end{equation}

\noindent{}where the summations are over the 8 EUVS Measurements, $E_{n,0}$ is the offset for the $n^{th}$  wavelength interval, and $P_i(t)$ and $Q_i(t)$ are the long- and short-term components of the $i^{th}$ EUVS Measurement, and $j_{i,n}$ and $k_{i,n}$ are the long- and short-term regression coefficients, respectively.  Note, the EXIS XRS measurements could also be included, but are not used in practice because their inclusion increases model uncertainty due to the variability at soft X-ray wavelengths typically being much larger than that at EUV wavelengths.  The long- and short-term components are defined by 

\begin{eqnarray}
P_i(t)=\frac{\left< X_i(t)\right>_T-X_{i,0} }{X_{i,0}}\\ \label{eqn_long}
Q_i(t)=\frac{X_i(t) - \left< X_i(t)\right>_T }{  \left< X_i(t)\right>_T } \label{eqn_short}
\end{eqnarray}

\noindent{}where $X_i(t)$ is the $i^{th}$ EUVS Measurement with reference offset, $X_{i,0}$, and (lagging) moving average, $\left< X_i(t)\right>_T$, over time, $T$.  The fractional units of $P_i(t)$ and $Q_i(t)$ reduces the sensitivity to subtle differences between spectral resolution and pre-flight calibrations of the historical model training data and the EUVS Measurements.

The short-term component is for modeling rapid solar flare induced variability whereas the long-term component is for modeling more gradual non-flaring variability.  The separation of flaring and non-flaring variability is done because different emission lines may contribute to the irradiance in a given spectral band during flares as a result of the significant temperature difference between flare and non-flare plasma. For example, the EUVS Measurement at 25.6 nm is primarily due to He II emissions during non-flaring times but dominated by Fe XIV emission during flares.  As such, $T$ is taken to be 6 hours in order to be significantly longer than most solar flares.  Specifically,  T is $\sim$4 times larger than the expected X-class flare duration as determined by \cite{veronig2002temporal}, who analyzed nearly 50,000 flares and found 90$\%$ of X-class flares  last less than 98 minutes.  Modeled irradiance for long duration flares, which can last for several hours, will be subject to increased error if the flare irradiance contributes measurably to the 6-hour average.

The offsets and coefficients in equations (\ref{eqn_model}) - (\ref{eqn_short}) are determined from historical data as follows:  $X_{i,0}$ is chosen as the minimum value for the historical data representing the $i^{th}$ EUVS Measurement.  $E_{n,0}$ and $j_{i,n}$ are found by multiple linear regression fitting of historical data representing $P_i(t)$ and the long-term average (nominally 6 hours) of $  E_n(t)$, $\left< E_n(t) \right>_T  $; and $k_{i,n}$ is found by multiple linear regression fitting of historical data representing  $Q_i(t)$ and ($E_n(t)-\left< E_n(t) \right>_T$).  These methods are described in further detail in Sections \ref{sec_long} and \ref{sec_short}, and the historical data are described in Section \ref{sec_data}.

The EUVS Model wavelength intervals are 5 nm wide from 5 to 115 nm (i.e. 5-10 nm, 10-15 nm, ..., 110-115 nm), with one 10 nm wide interval from 117 to 127 nm.  This is the official operational wavelength scheme produced in real time by NOAA and is the focus of this paper.  

The cadence of $X_i(t)$ (and, hence, the EUVS Model) is 30 seconds, where $X_i(t)$ is a 30-second centered running average of the $i^{th}$ EUVS Measurement.  The EUVS measurements all nominally have a 1-second cadence except for EUVS-C, which has a 3-second cadence.  $\left<X_i(t)\right>_T$ is a running average of the preceding 6 hours of $X_i(t)$.  The EUVS Model latency is 1 minute (i.e., EUVS Model irradiances are publicly available 1 minute after the corresponding EUVS Measurements are made).

The operational nature of the EUVS Model requires the continuous production of irradiances, even during times when some EUVS Measurements are unavailable.  This requires the derivation of independent sets of coefficients for the most likely contingencies. For brevity and clarity, we only focus on the nominal model, which assumes all EUVS measurements are available. The same methods described here are used to derive coefficients and model uncertainty for contingency cases, which will be reported on the web link given in the Acknowledgements Section.  The model uncertainty will generally be higher if the number of inputs is reduced, in particular, if there is no substitute measurement with a similar formation temperature available as discussed in Section \ref{sec_intro}. It is important to note that NOAA plans to maintain redundant satellites on orbit, reducing the likelihood that the EUVS Model will be operational at reduced performance for an extended period of time.

\section{Model Training Data and Cross-Calibration} \label{sec_data}

 \begin{figure}[h]
 \centering
 \includegraphics[scale=.5]{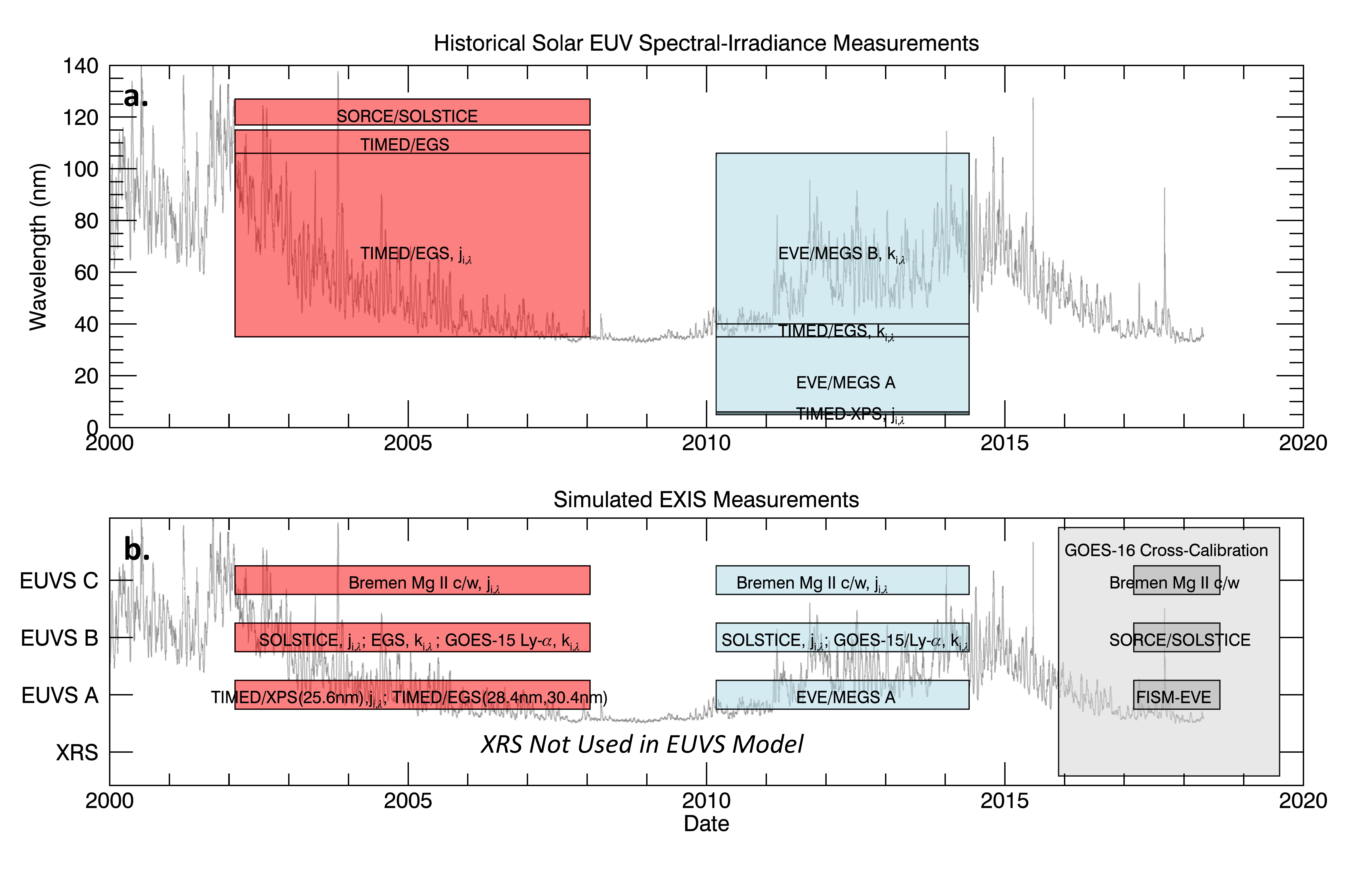}
 \caption{The time periods and wavelength ranges spanned by the (a) spectral irradiance and (b) EXIS measurement historical datasets used to determine EUVS Model coefficients.  Wavelengths or bands spanned are on the vertical axis and time is on the horizontal axis.  Table \ref{tab_datasets} provides details on the datasets used.} \label{fig_datasets}
  \end{figure}

\subsection{Historical Datasets}

The historical training datasets used to calculate the EUVS Model coefficients, and the time and spectral ranges over which they span are shown in Figure \ref{fig_datasets}, where Figure \ref{fig_datasets}a shows the spectral irradiance datasets used to create the simulated spectral intervals ($E_{n}$) and Figure \ref{fig_datasets}b shows the datasets used for the simulated EUVS Measurements ($X_i$).  Two separate time intervals of data are used to generate model coefficients, corresponding to periods when either TIMED/SEE or SDO/EVE are well calibrated. These intervals are distinguished in Figure 1 with red and blue shading.  Note, the TIMED/SEE dataset is well calibrated through 2011, but  the TIMED/SEE (red-shaded) interval stops at January 1, 2008 to prevent any sampling bias associated with including a disproportionate amount of data near solar minimum in the training dataset. Also shown in Figure  \ref{fig_datasets}b, with gray shading, is the time period and measurements used to cross-calibrate the historical training datasets with GOES-16.  The corresponding F10.7 flux is superimposed in the background of both panels for context.  Additionally, Figure \ref{fig_datasets} indicates cases where a dataset is only used for either the short or long -term regression coefficients.  For example, in the TIMED-EGS block in Figure \ref{fig_datasets}a the designation ``$j_{i,n}$" implies this dataset is only used to find the long-term regression coefficient, $j_{i,n}$.  Table \ref{tab_datasets} provides further information on the historical datasets, including data levels, versions, sampling and accuracy for the datasets shown in Figure \ref{fig_datasets}. The Purpose column in Table \ref{tab_datasets} indicates whether the dataset is used to find the short-term or long-term model regression coefficients, or for cross-calibrating the EXIS measurements with the historical datasets.

 \begin{table}[h]
 \scriptsize
 \caption{Historical training datasets used to determine the EUVS model coefficients. The Purpose column indicates if the data are used to compute the long-term and/or short-term model coefficients, or to cross-calibrate the historical data with the GOES-16 EUVS Measurements.}  \label{tab_datasets}
 \centering
 \begin{tabular}{l l l c l l l } 
 \hline
  Data Source & Purpose & Data Level & Version & Wavelength & Sampling & Accuracy \\
 \hline
  SORCE  / SOLSTICE $^a$ & Long-term, Cross-cal& 3 & 15 &117-141 nm& 0.1 nm & 5-8\%\\
  TIMED  / EGS $^b$ &Long and Short-term& 3 & 12 &26-117 nm& 1 nm & 5-20\%\\
  TIMED  / XPS $^c$ &Long-term& 3 & 12 &5-26 nm& 1 nm &20\%\\
  SDO  / MEGS $^d$ &Long and Short-term& 3 & 6 &6-106 nm& 0.1 nm & 1-7\%\\
      Bremen Composite Mg II Index$^e$ &Long-term, Cross-cal& Composite & 5 &275-285 nm& N/A & 0.3\%$^i$  \\
          FISM-EVE $^f$ &Cross-cal& Daily Average &1 &28-31 nm& 0.1 nm & 5-10\% \\
     GOES-15 /  EUVS E $^g$ &Short-term& Minute Average & 4 &121.6 nm& Lyman-$\alpha$ & 2\%$^h$ \\
    \hline
 \multicolumn{5}{l}{$^{a}$Solar Radiation and Climate Experiment / SOLar-STellar Irradiance Comparison Experiment} \\
 	\multicolumn{5}{l}{\citep{mcclintock2005solar}}\\
	\multicolumn{5}{l}{$^{b}$Thermosphere Ionosphere Mesosphere Energetics and Dynamics / EUV Grating Spectrograph}\\
	\multicolumn{5}{l}{ \citep{woods2005solar}}\\
	\multicolumn{5}{l}{$^{c}$ TIMED / XUV Photometer System \citep{woods2005solar}}\\
	\multicolumn{5}{l}{$^{d}$ Solar Dynamics Observatory / Multiple EUV Grating Spectrographs  \citep{woods2010extreme}}\\
	\multicolumn{5}{l}{$^{e}$ http://www.iup.uni-bremen.de/UVSAT/Datasets/mgii}\\
	\multicolumn{5}{l}{$^{f}$ FISM-EUV Variability Experiment \citep{chamberlin2018solar, thiemann2017maven} }\\
	\multicolumn{5}{l}{$^{g}$ https://www.ngdc.noaa.gov/stp/satellite/goes/doc/GOES$\_$NOP$\_$EUV$\_$readme.pdf} \\
	\multicolumn{5}{l}{$^{h}$ Uncertainty is with respect to scaling to SORCE/SOLSTICE absolute calibration.}\\
	\multicolumn{5}{l}{$^{i}$ \citep{woods2000improved}}
 \end{tabular}
 \end{table}

\subsection{EXIS Cross-Calibration with Historical Datasets}
The historical data used to represent the EXIS measurements (shown in Figure \ref{fig_datasets}b) are re-calibrated to best represent the GOES-16/EXIS measurements prior to finding the GOES model coefficients as follows:   For EUVS-A and EUVS-B, the  historical spectral irradiance data are interpolated to match the pixel scales of the GOES-16/EXIS detectors and then integrated over the same pixel masks used to compute the GOES-16/EXIS  Level 1b line irradiances.  These values are then cross-calibrated against the GOES-16/EXIS Level 1b line irradiances measured for the first 9 months of its mission using a first order Total Least Squares fit, which is implemented using the method of \cite{van1989extended}.  For EUVS-A, measurements from the EVE/MEGS-A channel do not overlap in time with GOES-16/EXIS.  In lieu of direct measurements for cross-calibrating EVE/MEGS-A with GOES-16, a version of the FISM model that is calibrated to EVE \citep{chamberlin2018solar} is used as an intermediary.  For EUVS-C, the Bremen Composite Mg II Index is fit against the GOES-16 Mg II Index native scale.  The Bremen Composite Mg II Index is sourced from the GOME-2B instrument onboard the MetOp-B satellite for the cross-calibration time period.  

\begin{figure}[h]
 \centering
 \includegraphics[scale=.4]{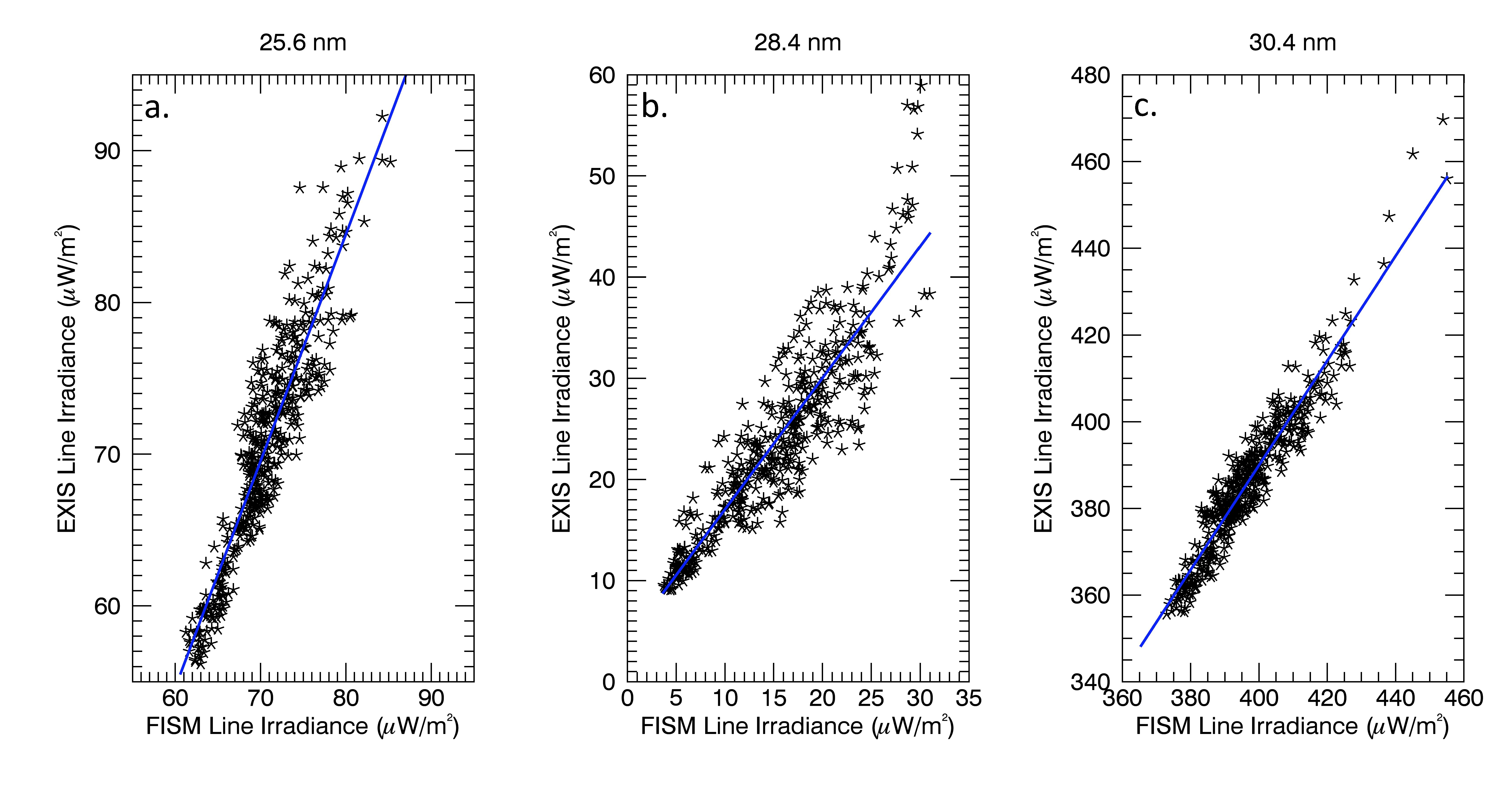}
 \caption{Scatterplots and best-fits for the EUVS-A line irradiances from FISM (a proxy for historical EVE data) and GOES-16/EXIS on the horizontal and vertical axes, respectively.  FISM is used in-lieu of EVE/MEGS-A because EVE/MEGS-A and GOES-16/EXIS did not make measurements contemporaneously.  Daily average data from 1 February 2017 through 10 April 2018 are used.} \label{fig_euvsa_scatter} 
  \end{figure}

\begin{figure}[h]
 \centering
 \includegraphics[scale=.6]{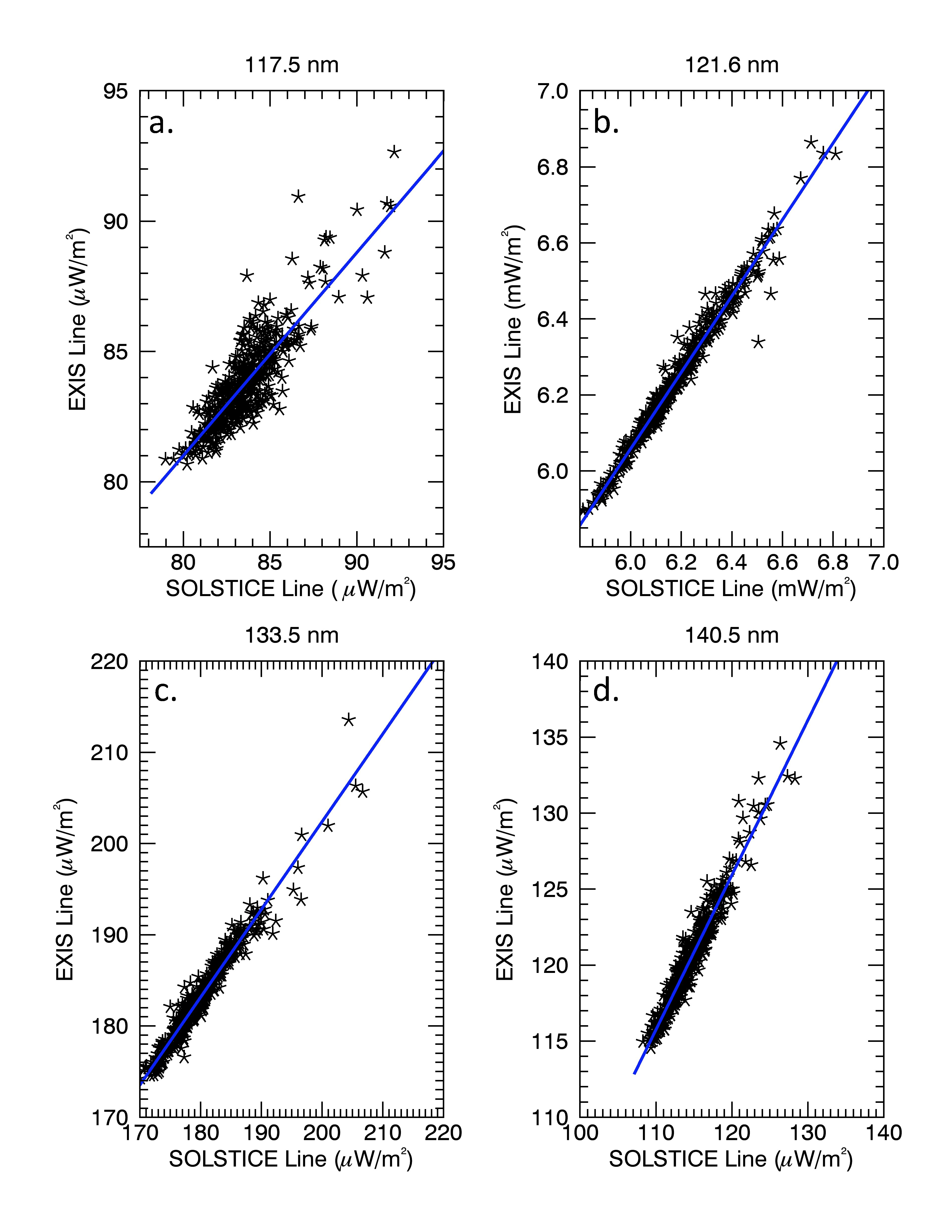}
 \caption{Scatterplots and best-fits of the GOES-16 EUVS-B line irradiances and SORCE/SOLSTICE  data. Daily average data from 1 February 2017 through 10 April 2018 are used.} \label{fig_euvsb_scatter} 
  \end{figure}

\begin{figure}[h]
 \centering
 \includegraphics[scale=.4]{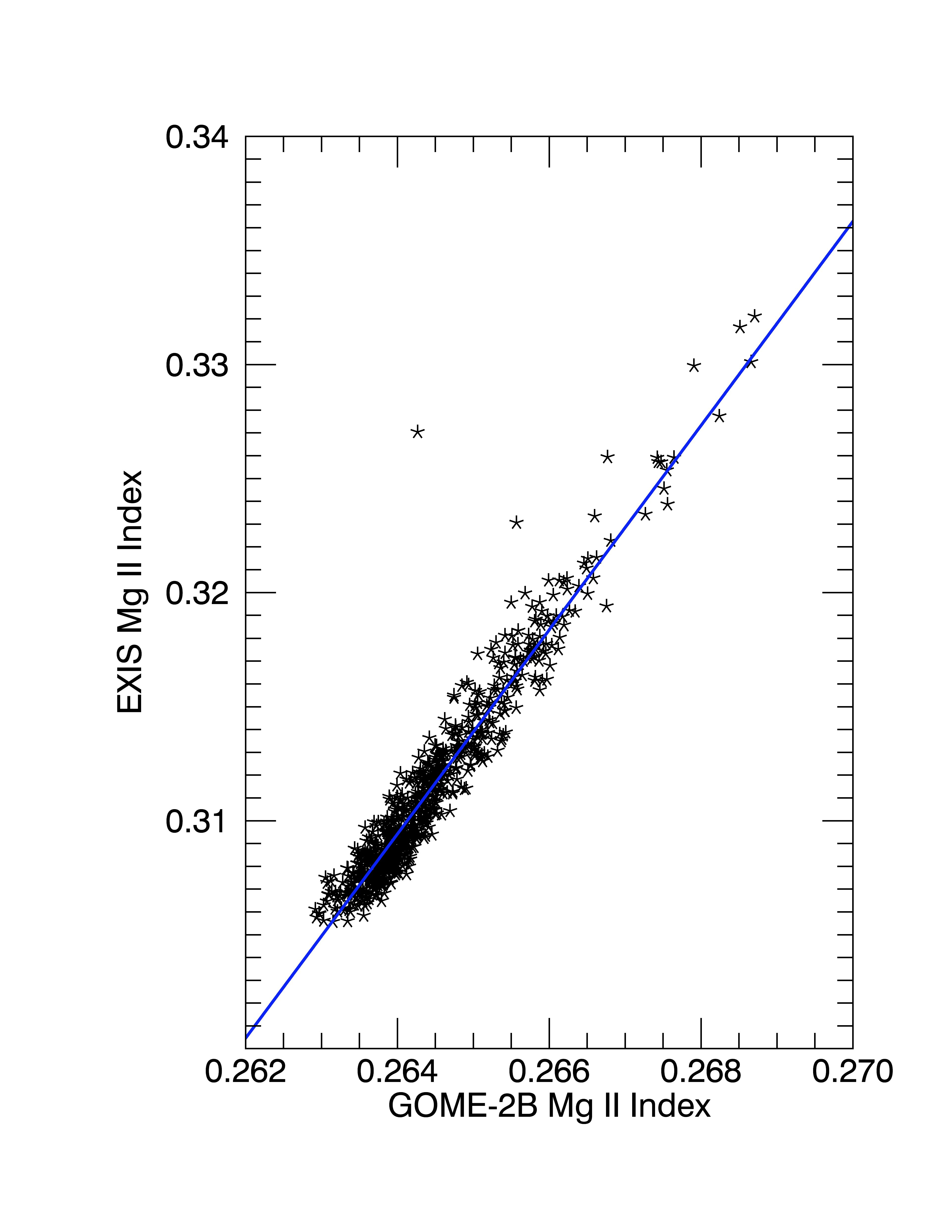}
 \caption{Scatterplot and best-fit of the GOES-16 EUVS-C Mg II Index and Bremen Composite Mg II Index. Daily average data from 1 February 2017 through 10 April 2018 are used.  The Bremen Composite Mg II Index source for this time period is from the GOME-2B instrument onboard the MetOp-B satellite.} \label{fig_euvsc_scatter} 
  \end{figure}

Figures \ref{fig_euvsa_scatter} - \ref{fig_euvsc_scatter}  show the cross-calibration data and fits for EUVS-A, B and  C, respectively.  The fit coefficients are given in Table \ref{tab_crosscal} along with the cross-calibration uncertainty, $\sigma_{CC,i}$. Note that $\sigma_{CC,i}$ only captures the uncertainty during the period of overlap, and uncorrected calibration drifts occurring in the training dataset, or future uncorrected drifts in the EUVS Measurements could increase the systematic error. These fits are used to simulate the EUVS Measurements using the historical training data.  For all channels except EUVS-A, $\sigma_{CC,i}$ is the uncertainty of the linear fit .  For EUVS-A lines used in the model intervals from 5-35 nm, $\sigma_{CC,i}$ is the quadrature sum of the cross-calibration uncertainty and the FISM model uncertainty, where the latter is reported in \citet{ thiemann2017maven}.  For EUVS-A lines used in the model intervals from 35-127 nm, an additional error term quantifying differences in the TIMED/SEE and SDO/EVE calibrations is included in the quadrature sum (in addition to the two aforementioned terms); this additional term is the uncertainty of the fit of the EUVS-A line bandpass measured by SDO/EVE to that measured by TIMED/SEE.  Therefore, two values are reported for $\sigma_{CC,i}$ for the EUVS-A lines in  Table \ref{tab_crosscal}; the value in parenthesis corresponds with model intervals from 35-127 nm whereas the other corresponds with model intervals from 5-35 nm.  There are also two values reported for the EUVS-B 121.6 nm line; the value in parenthesis is for the GOES-15/EUVS E channel cross-calibration, whereas the other value corresponds with SORCE/SOLSTICE.  

 \begin{table}[h]
 \caption{Cross-calibration fit coefficients and uncertainty for the eight EUVS Measurements.  The first two columns indicate the EUVS-Channel and line-center wavelength.  The slope and offset correspond with the line equation coefficients to estimate the EUVS measurements from the historical data shown in Figure \ref{fig_datasets}, and the corresponding uncertainty is shown in the right-most column.}  \label{tab_crosscal}
 \centering
 \begin{tabular}{l c c c c } 
 \hline
Channel & Wavelength (nm) & Slope & Offset & $\sigma_{CC,i}$ \\
 \hline
A & 25.6 & 1.49 & -35.02 $\mu$W & 4.1\% (6.2\%)\\
 A & 28.4 & 1.33 &  4.04 $\mu$W & 16.8\% (18.3\%)\\
A & 30.4 & 1.21 &  -92.40 $\mu$W & 1.3\% (3.5\%)\\
B & 117.5 & 0.80  & 17.37 $\mu$W & 1.1\%\\
B & 121.6 & 0.99 &  0.09 mW & 0.4\% (2.1\%) \\
B & 133.5 & 0.96 &  11.21 $\mu$W & 0.6\%\\
B & 140.5 & 1.03 & 2.65 $\mu$W & 0.7\%\\
C & 280 & 4.52 & -0.88 & 0.4\%\\
     \hline
 \end{tabular}
 \end{table}

\subsection{Model Uncertainties}

The available historical training datasets are partitioned roughly in half, with one-half  used to fit the model coefficients and the other half  used to compute the uncertainties. The uncertainty, $\sigma$, of a linear fit, say, $\hat{y}=mx_l+b$, between two linearly related, N-valued parameters $y_l$ and $x_i$ with fit coefficients, $m$ and $b$ is given by (e.g. \citet{taylor1997introduction})

\begin{equation}
\sigma=\sqrt{\frac{\sum\limits_{l=0}^N (y_l-\hat{y_l})^2}{N-2}}. \label{eqn_abs_sigma}
\end{equation}

\noindent Equation \ref{eqn_abs_sigma} is used in this study to compute the absolute uncertainties, with $\hat{y}_l$ corresponding with the $l^{th}$ model output and $y_l$ corresponding with the $l^{th}$ observation. In order to account for the cross-calibration uncertainty ($\sigma_{cc,i}$, listed in Table \ref{tab_crosscal}), gaussian noise with a standard deviation of $\sigma_{cc,i}$ is added to the $i^{th}$ model input prior to being used to compute $\hat{y}$. 

\noindent{}It is useful to compare uncertainties in percent units.  The percent uncertainty can be approximated as 

\begin{equation}
\sigma_{\%}\approx\frac{\sigma}{\left< y_l \right >}\times100\%,  \label{eqn_perc_sigma}
\end{equation}

\noindent{}An additional useful metric is the percent uncertainty relative to the typical variability as represented by the standard deviation,

\begin{equation}
\sigma_{s}\approx\frac{\sigma}{s}\times100\%  \label{eqn_perc_sigma}
\end{equation}

\noindent{}where $s$ is the standard deviation of the observations. It is important to keep in mind that $\sigma_{\%}$ and $\sigma_{s}$ are relative terms and, therefore, vary depending on the context in which they are computed. $\sigma_{ \%}$ will be larger (smaller) for lower (higher) irradiance values.  Similarly, $\sigma_{ s}$ will be larger (smaller) when computed over a weak (strong) solar cycle. $s$ and $\left< y_l \right >$ are computed over the time periods spanned by the historical data as shown in Figure \ref{fig_datasets}.

In this study, the Bootstrap method \citep{efron1979bootstrap} is used to estimate the model uncertainty. 1000 synthetic datasets are generated by resampling (with replacement) the available $[y,\hat{y}]$ pairs of values, and values for $\sigma$ are computed from each synthetic dataset. It is found that the standard deviation of $\sigma$ is typically $\sim$3$\%$ (i.e. a reported 10$\%$ uncertainty has a spread of $\sim$0.3$\%$) for the long-term model component and $\sim$15$\%$ for the short term model component. $\sigma$ values and their standard deviations as determined by Bootstrapping are reported in Sections \ref{sec_long} and \ref{sec_short}.

\section{6-Hour Average (Long-Term) Model Component}  \label{sec_long}

\subsection{Long-Term Model Formulation} \label{sec_long_form}

The first two terms on the right-hand side of Equation \ref{eqn_model} comprise the long-term model component.  The historical spectral irradiance data are interpolated to 1-nm sampling for determining the model coefficients, $E_{n,0}$ and $j_{i,n}$. The regression coefficients are found using the Interactive Data Language (IDL) multiple linear regression program, regress.pro, via an iterative scheme: Regressors with negative coefficients or that contribute to less than 5$\%$ of the fit (as a result of having relatively small coefficients) are set to zero, and the fits are re-computed until all coefficients are positive and all regressors contribute to at least 5$\%$ of the fit.  It was found that for particularly noisy intervals, multiple linear regression can result in model error that is larger than that from single linear regression using a single EUVS Measurement.  To account for this, a multiple linear regression model using the combined eight EUVS Measurements and eight single linear regression models using the eight EUVS Measurements individually are computed.  The corresponding model errors, defined as the quadrature sum of the mean and standard deviation of the model-measurement difference, are compared, and the model with the smallest error is selected for each 1-nm interval.  

The coefficients for the 1-nm intervals corresponding with a given EUVS Model interval are then summed over the wavelength range defined by the interval. An example is given to demonstrate how the fit coefficients determined from the 1 nm wide intervals are combined to correspond with a 5 nm wide interval: The coefficients for the 5-10 nm interval, $E_{5-10,0}$ and $j_{i,5-10}$ are found from the five sets of 1-nm sampling coefficients as follows:
\begin{eqnarray}
E_{5-10,0}=\frac{1}{5}\sum_{\lambda = 5 nm}^{9 nm}E_{\lambda,0}\\
j_{i,5-10}=\frac{1}{5}\sum_{\lambda = 5 nm}^{9 nm}j_{i,\lambda}
\end{eqnarray}
The factor of $\frac{1}{5}$ rescales the coefficients for spectral irradiance from 1-nm to 5-nm intervals , with units of irradiance per nm.  For the 117-127 nm interval, the coefficients are scaled by $\frac{1}{10}$.

\subsection{Long-Term Model Component Results}\label{sec_lt_results}

In order to more accurately characterize the model uncertainty, the historical datasets are partitioned approximately in half, with the half corresponding with  lower solar activity being used to compute the model coefficients, while the half corresponding with higher solar activity is used for model-measurement comparison to quantify the model uncertainty. The long-term component \emph{model} uncertainties, $\sigma_{LTM}$ , for each EUVS Model interval are reported in Table \ref{tab_model_long} in absolute units, units of percent uncertainty and percent variability. Additionally, Table \ref{tab_model_long} reports the standard deviation of the uncertainties, $s_{\sigma }$, in absolute units and the long-term component \emph{instrument} percent uncertainties, $\sigma_{LTI}$, the latter of which correspond with the signal-weighted average instrument accuracies of the historical spectral-irradiance measurements that are used to train the EUVS Model.  

$\sigma_{LTM}$ is the random uncertainty (i.e. the classical precision) of the EUVS Model long-term component, whereas $\sigma_{LTI}$ is the Model's systematic uncertainty (i.e. the classical accuracy). In other words, $\sigma_{LTI}$ characterizes the magnitude of a constant bias in the model estimates, while $\sigma_{LTM}$ characterizes the error in the 6-hour variability estimated by the model. It is important to note the model absolute accuracy strictly depends on the absolute accuracy of the historical training data, and for a number of wavelength intervals, the random uncertainty ( $\sigma_{LTM}$) is smaller than the uncertainty of the systematic bias ( $\sigma_{LTI}$).

The sensitivity of the model parameters to the model training period used is investigated by exchanging the model training and validation periods. When the model is trained using the dataset half corresponding with higher solar activity and validated with the less active half, $\sigma_{LTM}$ is smaller for all intervals except three (the 50-55 nm, 55-60 nm, and 75-80 nm intervals). The average difference in $\sigma_{LTM}$ is 15$\%$. Additionally, the model coefficients found from the two different training periods tend to be different, and the sets of optimal model inputs for a model interval also tend to differ. This suggests that model error can be reduced by using training data from periods that have solar activity levels comparable to that of the period for which the model is being applied.

\begin{figure}
 \centering
 \includegraphics[scale=.35]{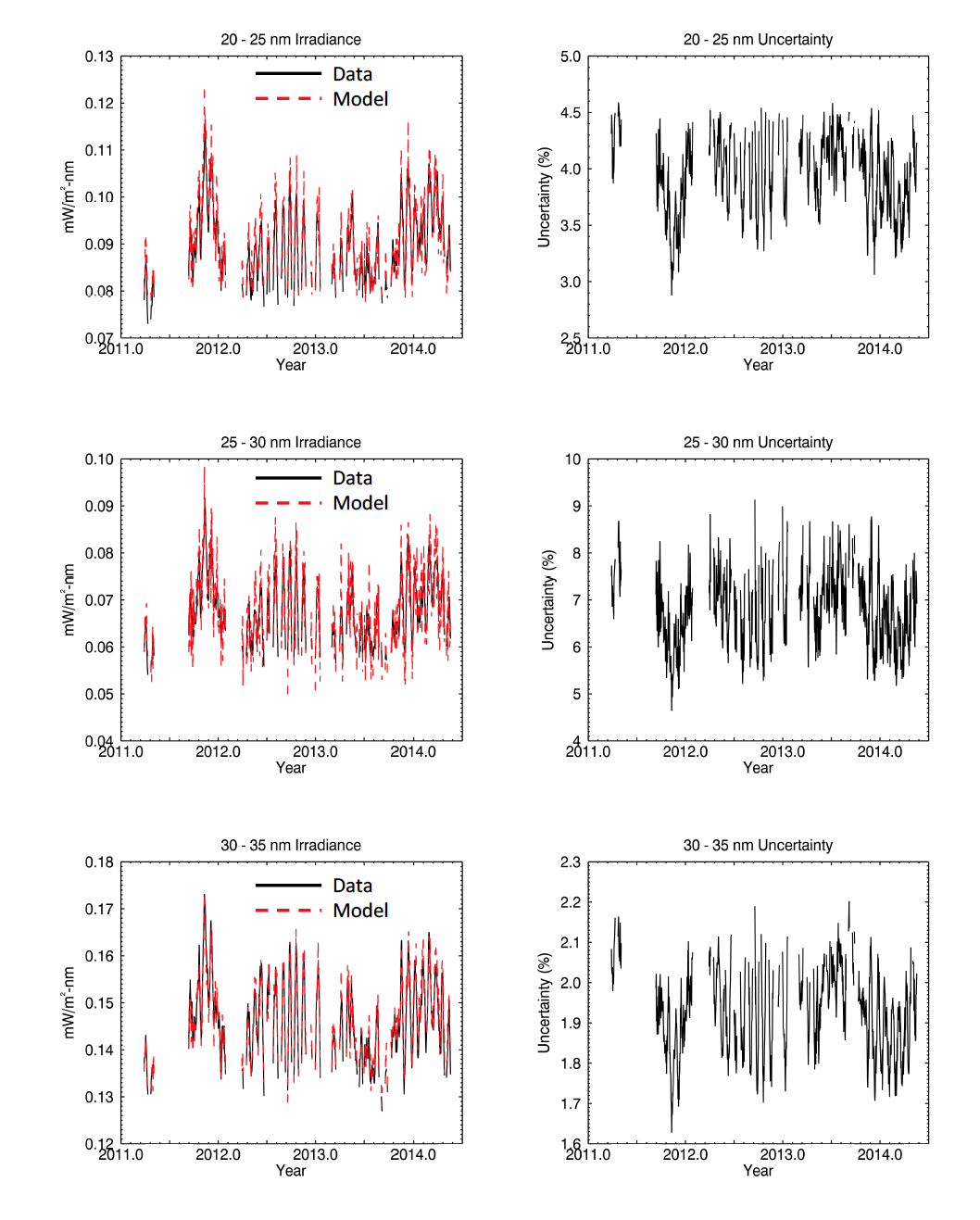}
 \caption{Sample model-measurement comparisons for intervals derived from SDO/EVE data. } \label{fig_mod_meas_1} 
  \end{figure}
\begin{figure}
 \centering
 \includegraphics[scale=.35]{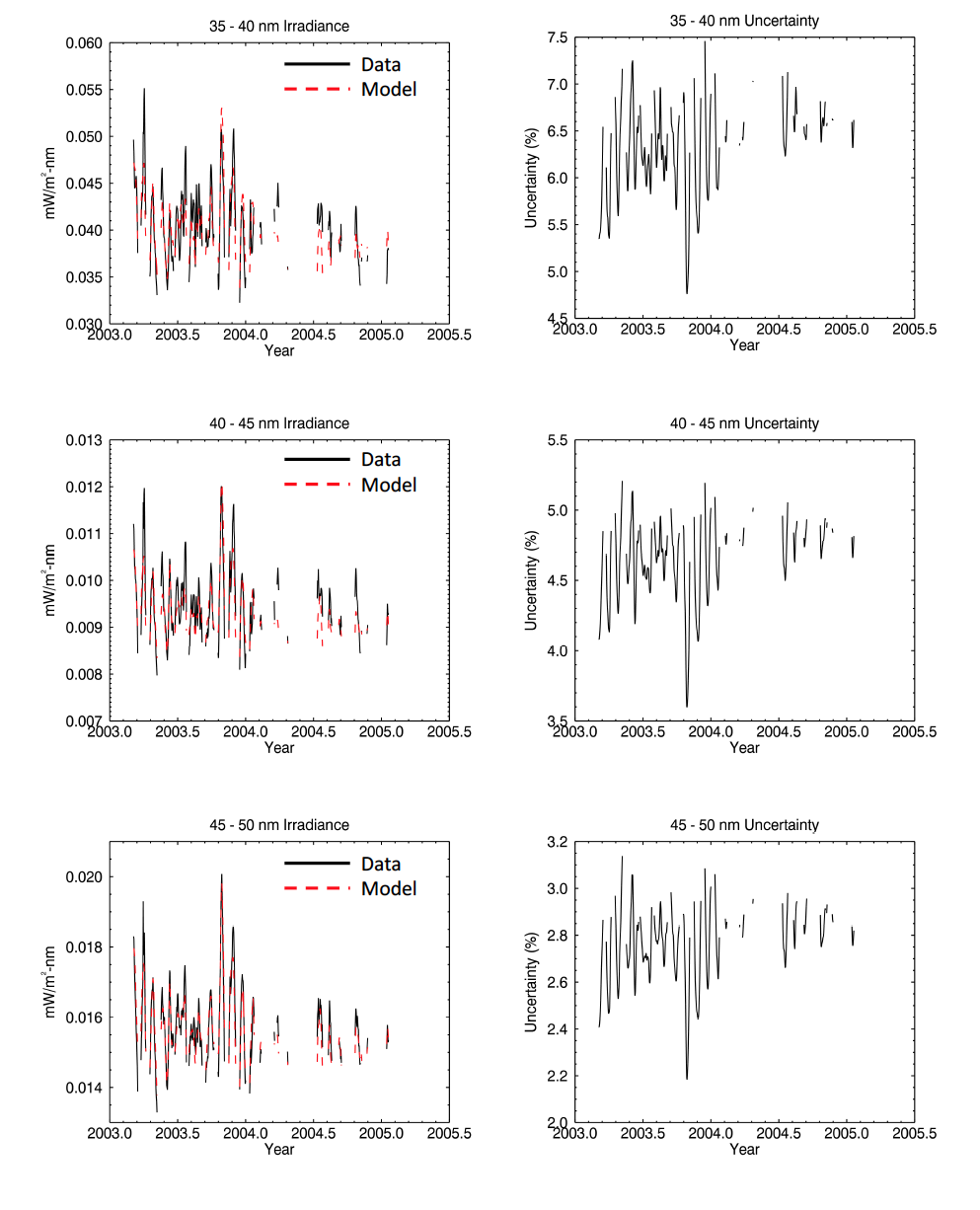}
 \caption{Sample model-measurement comparisons for intervals derived from TIMED/EGS data. } \label{fig_mod_meas_2} 
  \end{figure}

Sample model measurement comparisons for 6 intervals are shown in Figures \ref{fig_mod_meas_1} and \ref{fig_mod_meas_2} for model intervals derived from SDO/EVE and TIMED/EGS data, respectively.  The modeled and measured irradiances are shown in the left-hand columns with red and black curves, respectively.  The right-hand columns show time series of the model uncertainty in percent units.  Here, the percent uncertainty is defined as $\sigma_{LT}$ divided by the daily modeled irradiance multiplied by 100$\%$. The training and validation time periods can be inferred from Figures 5 and 6: Times when data are shown correspond with the validation dataset times; times when data are absent between 30 April 2010 and 26 May 2014 (8 February 2002 and 1 January 2008) correspond with the training dataset times for the SDO/EVE (TIMED/EGS) data interval.

 \begin{table}[h]
  \scriptsize
   \caption{Long-term model uncertainty in absolute units along with its standard deviation. Uncertainties are also reported in percent units, both relative to the mean and observed variability (as measured by the standard deviation).}  \label{tab_model_long}
 \centering
 \begin{tabular}{l c c c c c} 
 \hline
 Model Interval (nm) & $\sigma_{LTM}$ $(\mu W/m^2$-nm)&$s_{\sigma}$ $(\mu W/m^2$-nm)& $\sigma_{LTM}/<E_{n,L}>$ ($\%$) & $\sigma_{LTM}/s_{n,L}$ ($\%$) &  $\sigma_{LTI} ($\%$)$\\
 \hline
5-10 & 3.095 & 0.078 & 5.5 & 24.1 & 2.2 \\ 
10-15 & 1.064 & 0.037 & 5.3 & 30.8 & 2.2 \\ 
15-20 & 4.817 & 0.122 & 4.3 & 27.6 & 1.1 \\ 
20-25 & 3.375 & 0.089 & 4.4 & 20.5 & 1.5 \\ 
25-30 & 4.475 & 0.122 & 8.0 & 32.0 & 2.9 \\ 
30-35 & 2.737 & 0.069 & 2.0 & 18.1 & 3.0 \\ 
35-40 & 2.539 & 0.102 & 9.0 & 39.3 & 10.6 \\ 
40-45 & 0.430 & 0.021 & 5.8 & 40.3 & 10.1 \\ 
45-50 & 0.429 & 0.018 & 3.4 & 27.2 & 10.4 \\ 
50-55 & 0.485 & 0.025 & 5.2 & 29.5 & 10.8 \\ 
55-60 & 0.715 & 0.033 & 3.7 & 42.7 & 13.7 \\ 
60-65 & 0.925 & 0.045 & 4.4 & 43.0 & 19.1 \\ 
65-70 & 0.139 & 0.006 & 2.8 & 32.9 & 9.3 \\ 
70-75 & 0.177 & 0.008 & 3.0 & 41.4 & 9.5 \\ 
75-80 & 0.483 & 0.026 & 3.4 & 66.0 & 13.5 \\ 
80-85 & 0.598 & 0.028 & 3.3 & 39.8 & 12.8 \\ 
85-90 & 1.244 & 0.058 & 3.7 & 34.2 & 12.9 \\ 
90-95 & 1.030 & 0.048 & 3.6 & 33.2 & 12.7 \\ 
95-100 & 1.534 & 0.054 & 4.4 & 44.8 & 18.1 \\ 
100-105 & 1.870 & 0.078 & 3.9 & 35.2 & 18.8 \\ 
105-110 & 0.439 & 0.015 & 2.3 & 24.5 & 9.5 \\ 
110-115 & 0.500 & 0.019 & 2.5 & 31.9 & 12.4 \\ 
117-127 & 2.223 & 0.025 & 0.3 & 2.9 & 10.0 \\ \hline
 \end{tabular}
 \end{table}

\section{Minute-Average (Short-Term) Model Component}  \label{sec_short}

\subsection{Short-Term Model Formulation}
The short-term model component is the second term on the right hand side of Equation \ref{eqn_model}.  The model coefficients are found using multiple linear regression at EUVS Model resolution rather than 1 nm sampling to improve the signal-to-noise ratio (SNR) of short time scale variability observations.  Because the largest variations in solar EUV variability at short time-scales are due to solar flares, the regression fitting is done using the peak irradiance enhancements occurring during solar flares.   Typical fits result in a small constant term, which is collected  in $E_{n,0}$ in Equation \ref{eqn_model} along with the constant term from the long-term component.  The SDO/MEGS spectral irradiances are used for the 6-105 nm wavelength range, except for the 35-40 nm interval where the TIMED/EGS spectral irradiances are used.  TIMED/EGS spectral irradiances are also used for the 105-115 nm wavelength range, and the SORCE/SOLSTICE irradiance is used for the 117-127 nm wavelength interval, each of which will be discussed independently in the following paragraphs.

The EVE/MEGS dataset is the most extensive dataset measured to date of EUV variability at short time-scales, having measured EUV variability at 0.1 Hz for thousands of flares.  MEGS A made measurements near continuously, while MEGS B has a lower duty cycle for degradation mitigation, resulting in MEGS B having made fewer flare observations.  For every M-class or larger flare observed by MEGS A, the pre-flare background and flare peak irradiance values are identified for each 1 nm interval in the MEGS wavelength range.  This is done by manually identifying the time range corresponding with the background and peak irradiances using the bright flaring 13.29 Fe XXIII line and then using an automated program to average the background and identify the peak values.  This same process was used by \citet{thiemann2018center}, and the reader is referred to that paper for further details.  

\begin{figure}[h]
 \centering
 \includegraphics[scale=.4]{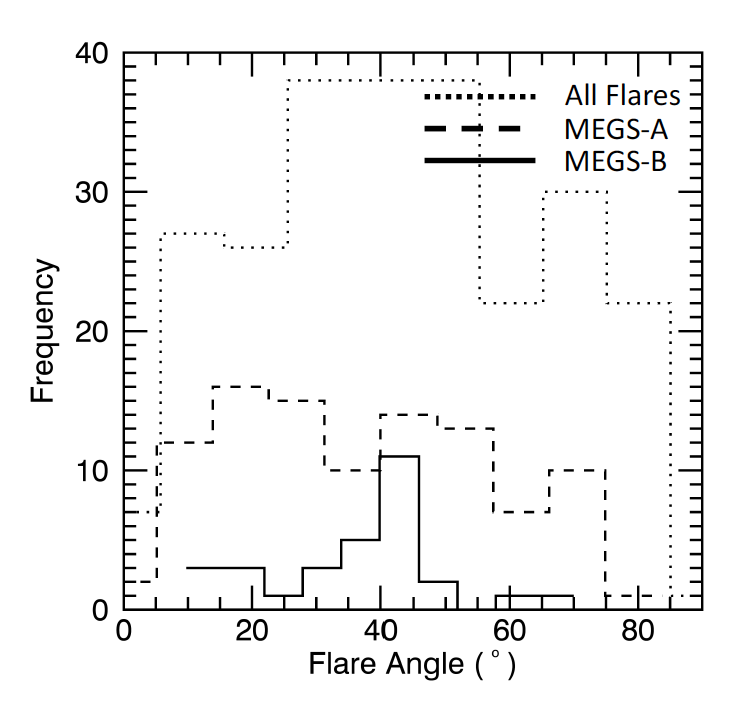}
 \caption{Histograms of all M class flares observed by MEGS A (dotted), flares used to to compute model coefficients in the MEGS-A (dashed) and MEGS-B (solid) wavelength ranges. } \label{fig_flare_ang} 
  \end{figure}

A total of 249 (50) flares are identified in the MEGS A (B) wavelength range.  Of these flares, the 100 flares with the largest enhancement at 13.5 nm are used to compute model coefficients in the MEGS A wavelength range, while 30 flares with the largest enhancement at 97.5 nm are used for finding model coefficients in the MEGS B wavelength range.  The MEGS B sample size is limited to 30 because including more flares increases the model error across many intervals as a result of having a low SNR.  However, the fit uncertainties approach an asymptotic value after a sample size of $\sim$10, so 30 flares is a large enough sample size to accurately characterize the model error.  13.5 nm and 97.5 nm are used for bright flare identification because they both show significant enhancement during flares, and hence, have high SNR.  The 13.5 nm interval is dominated by a coronal Fe XXIII line during flares, and hence has relatively small opacity during flares.  Although, it is important to note that even hot coronal lines, typically considered to be optically thin, limb darken during flares \citep{thiemann2018center}.  On the other hand, the 97.5 nm interval in the MEGS B wavelength range is dominated by C III and Lyman-$\gamma$, the latter of which strongly limb darkens due to the high abundance of H in the Sun's atmosphere.  This inherently biases the sample set of flares in the MEGS B wavelength range to those having originated away from the limb.  As an alternative, the 49.5 nm interval, which is dominated by the hot coronal and, hence, less opaque Si XII emission, was also considered for flare identification.  However, it was found that the SNR of the training set greatly decreased for many intervals in the MEGS B range as a result of their being dominated by more optically thick transition region emissions.  Therefore, it was decided to use the 97.5 nm interval in order to have higher model accuracy for flares located nearer to disk center, as they tend to be more geo-effective \citep{qian2010flare} at a cost of higher model uncertainty for flares located nearer to the limb. The full set of flares, which includes both disk-center and limb flares, is used to compute the model uncertainty. As such, the reported uncertainties are consistent with the average flare, and likely over-estimated (under-estimated) for disk-center (limb) flares. Figure \ref{fig_flare_ang} shows a histogram of the flare disk-location for all 249 flares considered as well as those used in the MEGS A and MEGS B wavelength ranges.  Note, in both cases, flares at 80$^o$ or further from disk center were discarded to avoid using flares that may be partially occulted by the limb.

The MEGS dataset only contains the EUVS-A lines, and there are no corresponding high time-cadence measurements made for the EUVS-B and EUVS-C lines except for the 121.6 nm Lyman-$\alpha$ line, which is measured by the GOES-15/EUVSE instrument.  As a result, only simulated EXIS measurements for the EUVS-A lines and Lyman-$\alpha$ are available for generating the model coefficients for the intervals ranging from 5 nm to 105 nm (except the 35-40 nm interval).  

The model intervals ranging from 105 nm to 115 nm are derived from the TIMED-SEE dataset.  The 35 to 40 nm interval is also derived from the TIMED-SEE dataset because the MEGS-B flare measurements had very low SNR over this wavelength range.  Flares of magnitude M5 or larger occurring between 30 October 2002 and 18 October 2008 during a SEE observation are used to derive the model coefficients (where the flare timing is taken to be that reported in the GOES events list returned by the pr$\_$gev.pro SolarSoft routine).  This results in a training dataset of 36 flares.

The  117-127 nm interval is derived from the SORCE/SOLSTICE dataset using all X1 or larger flares that occurred during SOLSTICE observations between mid-2003 and 2011, resulting in 19 flares.  Because SOLSTICE does not measure short-ward of 117 nm, only simulated EUVS-B lines are used to generate the model coefficients for this wavelength interval.



\subsection{Short-Term Model Component Results}

Table \ref{tab_model_short} shows the corresponding short-term model uncertainty ($\sigma_{STM}$) values in absolute units and their standard deviations (as determined by Bootstrapping). Values for for $\sigma_{STM}$ are also given in percent units relative to the mean and standard deviation of the flare peak intensities of the training dataset. The relative magnitude of the mean flare peak values to the peak daily average values are given in the right-most column in percent units for context. As is the case with the long-term model component, $\sigma_{STM}$ is the random uncertainty, and the short-term instrument uncertainty ($\sigma_{STI}$, not shown) is the systematic uncertainty of the short-term model component.
\noindent{}The values for $\sigma_{STI}$ are similar to $\sigma_{LTI}$, but somewhat smaller from 35-105 nm due to the smaller absolute uncertainty of SDO/EVE relative to TIMED/SEE over these wavelengths.

\begin{figure}[h]
 \centering
 \includegraphics[scale=.65]{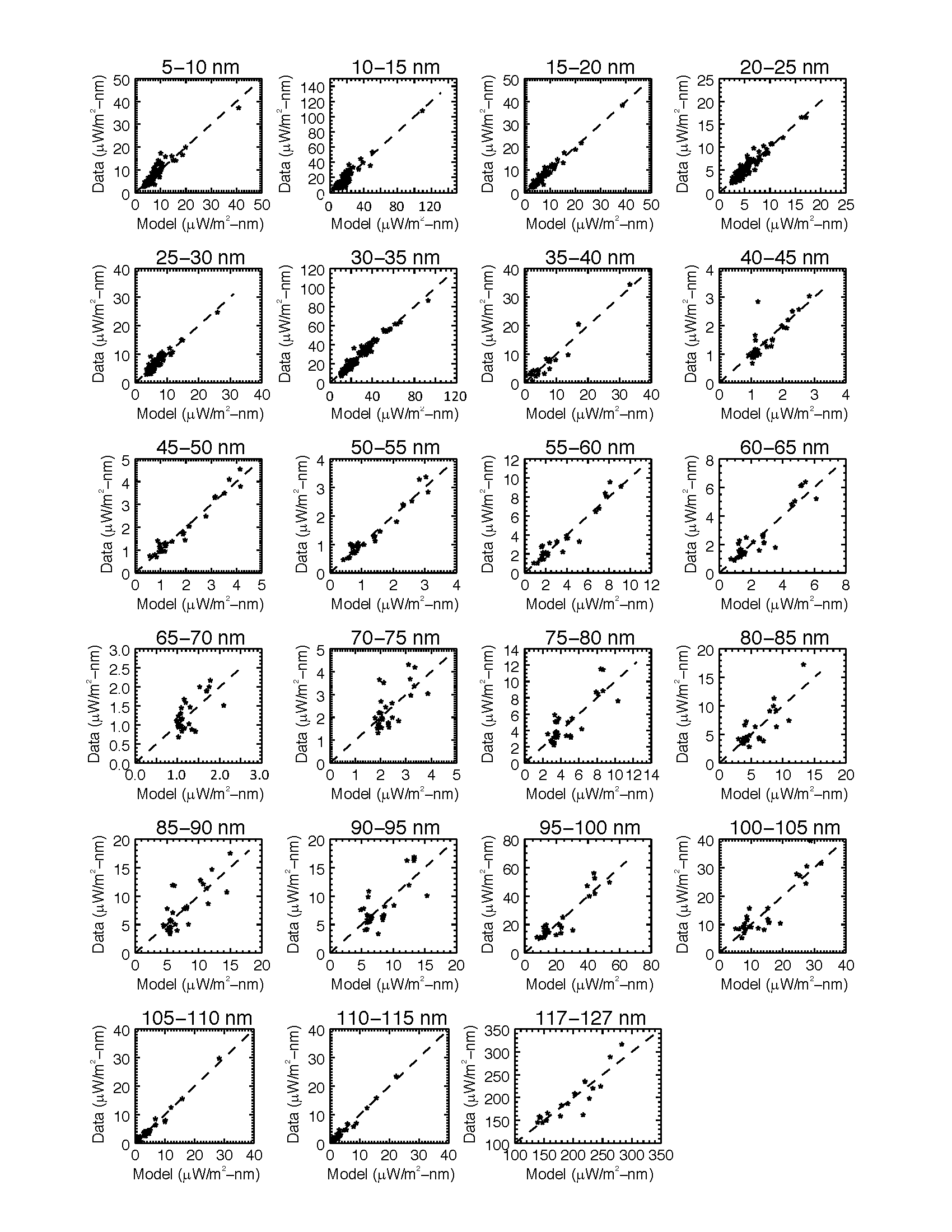}
 \caption{Scatterplots of measured and modeled peak flare irradiances for the flares used to calibrate the EUVS Model short-term component. } \label{fig_flare_scatter} 
  \end{figure}

Figure \ref{fig_flare_scatter} shows scatter-plots of the measured and modeled peak flare enhancements from the training data for all 23 EUVS model intervals.  From Figure \ref{fig_flare_scatter} and Table \ref{tab_model_short}, it is evident that the EUVS model predicted peak flare enhancements are highly correlated with measurements for most intervals.  

\begin{figure}
 \centering
 \includegraphics[scale=.65]{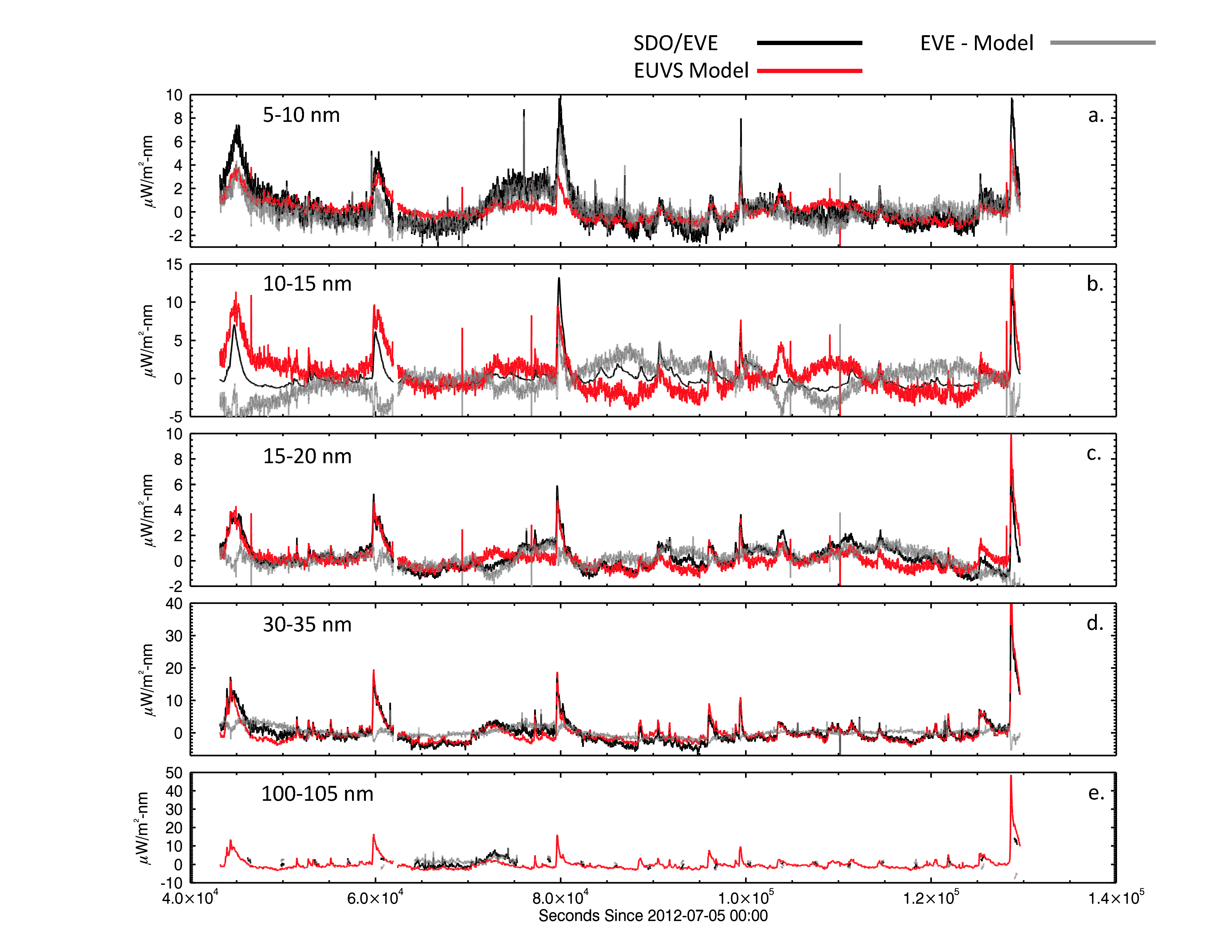}
 \caption{A comparison of measured (black) and modeled (red) short-term variability for a 24 hour period beginning on 5 July 2012 12:00 UT. Measurement-Model differences are shown in gray.   Five intervals are shown that were selected because they typically show relatively large enhancements during flares; the interval wavelength range is given on each panel.} \label{fig_flare_example_2012} 
  \end{figure}

\begin{figure}
 \centering
 \includegraphics[scale=.65]{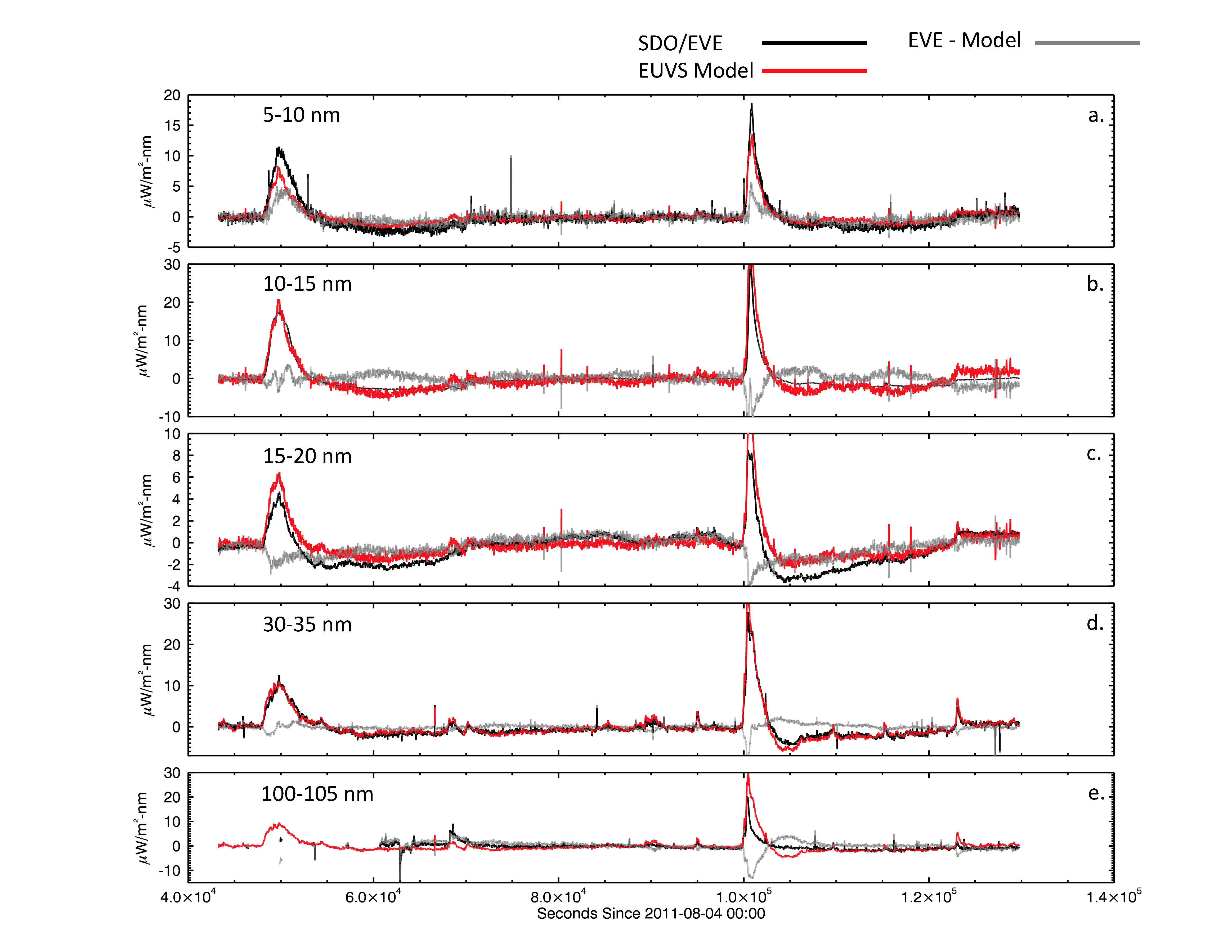}
 \caption{Same as Figure \ref{fig_flare_example_2012}, but for a period beginning on 4 August 2011 12:00 UT. } \label{fig_flare_example_2011} 
  \end{figure}

In Figures \ref{fig_flare_example_2012} and \ref{fig_flare_example_2011}, the short-term EUVS Model predictions are compared with SDO/EVE measurements for two sample days, beginning  at 12:00 UT on 5 July 2012 and 4 August 2011, respectively.  Each figure shows measured and modeled short-term irradiances for a 24-hour period using black and red curves, respectively, for five wavelength intervals chosen because they tend to show a relatively large flare enhancement. The curves represent the short-term solar variability, i.e. the right-most term in Equation \ref{eqn_model}. Measurement-model differences are shown in each panel with gray curves.   The SDO/EVE and simulated EXIS measurement data have the 6-hour moving average removed and are plotted at 10 second cadence with a 30-second ( 3-sample) moving average applied.  The missing SDO/EVE data in Panels e of both figures is because SDO/EVE reduces its duty cycle at this wavelength range for degradation mitigation.  

Considering first the non-flaring short-term variability that is most prevalent in Figure \ref{fig_flare_example_2012}, the EUVS Model captures this variability well in the 5-10 nm and the 30-35 nm wavelength intervals, but to a lesser degree in the 10-15 nm and 15-20 nm intervals.  The EUVS Model is expected to model variability in the 30-35 nm interval well because it is dominated by the 30.4 nm He II emission and this emission is used as a direct input to the EUVS Model.  The emissions forming from 5-20 nm are formed primarily at hot temperatures in the solar corona.  Therefore, their corresponding emission measures can be invisible to the EXIS measured emissions, which are formed at relatively cooler temperatures.  The short-term variability in these intervals is modeled predominantly using the 25.6 nm and 28.4 nm EXIS Measurements.  The reason for the relatively large magnitude of the 25.6 nm measurement coefficient for these wavelength regions dominated by coronal forming emissions is because  the 25.6 nm EXIS measurement is dominated by an Fe XXIV emission at 25.53 nm during solar flares.  As such, the flare enhancements are well modeled in the 10-20 nm intervals, when the Fe XXIV emission dominates the EXIS measurement, but the model performance decreases during non-flaring periods when the cooler forming He II emission dominates.

Figure \ref{fig_flare_scatter} quantifies how well the EUVS Model predicts the peak flare emission.  From Figure \ref{fig_flare_scatter}, there appears to be some slight non-linearity between the model predictions and measurements.  For example, in the 5-10 nm panel, the slope is steeper below 10 $\mu$W/m$^2$ than it is above 10 $\mu$W/m$^2$, resulting in the model under-predicting enhancements for (less frequent) larger flares.  

Figures \ref{fig_flare_example_2012} and \ref{fig_flare_example_2011} provide insight into how well the EUVS Model captures the time evolution of solar flare emissions.  Flare emissions at EUV wavelengths can pass through a number of phases as they evolve in time \citep{woods2011new}, including an initial impulsive phase followed by a gradual phase, which can be delayed in time depending of the emission formation temperature \citep{thiemann2017time}; later flare phases typically associated with eruptive flares include a dimming phase and a late phase.  Examples of all these phases except the late phase can be seen in Figures \ref{fig_flare_example_2012} and \ref{fig_flare_example_2011}.  Three impulsive flares are evident in Figure \ref{fig_flare_example_2012}, occurring near $6\times 10^4 $, $8\times 10^4 $ and $13\times 10^4 $ seconds.  These are all well-modeled in the 30-35 nm interval, again due to it being driven by the 30.4 nm EXIS measurement.  The impulsive phase for these flares is modeled poorly in the 15-20 nm interval, where it is under-predicted for the first two flares and over-predicted for the third flare.  This is likely because this interval is driven predominantly by the  Fe XXIV 25.53 nm emission during flares, which typically lacks an impulsive phase.  The model does not adequately capture the gradual phase delay in the 10-20 nm intervals for the first two flares in Figure \ref{fig_flare_example_2012}.  In the 10-15 nm interval, the model over-predicts the gradual phase delay, whereas in the 15-20 nm interval the gradual phase delay is slightly under predicted, for both flares.  Coronal dimming occurs after both flares in Figure \ref{fig_flare_example_2011} and is responsible for the emission decrease in panels c and d following one or both flares.  The apparent dimming in Panels a and b is a result of subtracting off the lagging 6-hour average from the data and the relatively long duration of the two flares.  The dimming is under-predicted in the 15-20 nm range for both flares because the Fe XXIV emission typically does not dim.  For the second flare, dimming appears in the model but not in the data because this interval is driven by the 30.4 nm EXIS measurement, which showed dimming for this flare.

\begin{table}[h]
  \scriptsize
   \caption{Short-term model uncertainty in absolute units along with its standard deviation. Uncertainties are also reported in percent units, both relative to the mean and observed flare peak intensity of the training data set. The right-most column shows the mean flare peak intensity relative to the daily average in percent units for context.}  \label{tab_model_short}
 \centering
 \begin{tabular}{l c c c c c} 
 \hline
 Model Interval (nm) & $\sigma_{STM}$ $(\mu W/m^2$-nm)&$s_{\sigma}$ $(\mu W/m^2$-nm)& $\sigma_{LTM}/<E_{n,S}>$ ($\%$) & $\sigma_{LTM}/s_{n,S}$ ($\%$) &  $<E_{n,S}>/<E_{n,L}> ($\%$)$\\
 \hline
5-10 & 1.739 & 0.197 & 23.6 & 35.9 & 13.2 \\ 
10-15 & 4.138 & 0.407 & 26.8 & 31.1 & 77.3 \\ 
15-20 & 0.991 & 0.104 & 14.8 & 20.0 & 6.0 \\ 
20-25 & 0.844 & 0.077 & 15.8 & 30.0 & 7.0 \\ 
25-30 & 1.060 & 0.104 & 16.8 & 34.1 & 11.2 \\ 
30-35 & 3.297 & 0.347 & 12.7 & 21.7 & 19.4 \\ 
35-40 & 1.840 & 0.252 & 27.0 & 48.5 & 24.2 \\ 
40-45 & 0.347 & 0.121 & 23.3 & 55.5 & 20.2 \\ 
45-50 & 0.230 & 0.032 & 12.8 & 20.8 & 14.2 \\ 
50-55 & 0.186 & 0.026 & 14.0 & 22.6 & 14.2 \\ 
55-60 & 0.682 & 0.116 & 19.2 & 27.4 & 18.4 \\ 
60-65 & 0.649 & 0.090 & 26.2 & 38.4 & 11.7 \\ 
65-70 & 0.303 & 0.033 & 24.2 & 72.8 & 25.5 \\ 
70-75 & 0.656 & 0.095 & 27.9 & 77.1 & 40.3 \\ 
75-80 & 1.502 & 0.169 & 30.8 & 57.7 & 34.2 \\ 
80-85 & 1.765 & 0.227 & 30.9 & 57.2 & 31.8 \\ 
85-90 & 2.362 & 0.370 & 31.3 & 63.1 & 22.7 \\ 
90-95 & 2.513 & 0.278 & 31.0 & 65.3 & 28.4 \\ 
95-100 & 5.459 & 0.807 & 25.0 & 39.3 & 62.1 \\ 
100-105 & 3.972 & 0.598 & 28.5 & 45.3 & 29.4 \\ 
105-110 & 1.070 & 0.140 & 19.6 & 17.2 & 28.4 \\ 
110-115 & 0.943 & 0.123 & 20.0 & 17.8 & 23.3 \\ 
117-127 & 21.593 & 4.246 & 11.0 & 43.6 & 26.5 \\
\hline
 \end{tabular}
 \end{table}

\section{Discussion} \label{sec_disc}

All coefficients needed to compute EUVS Model irradiances from the EUVS Measurements are reported in the Appendix. The EUVS model coefficients determined here from irradiance measurements made during Solar Cycles 23 and 24 are expected to apply to future solar cycles because the relationship between EUV emissions and the thermal structure of the solar atmosphere are expected to be invariant between solar cycles. Because the less (solar) active half of the available data were used to determine the model coefficients and the more active half of the data were used to characterize its uncertainty, the model performance has been validated for periods of higher activity than that occurring during the period used to compute the model coefficients. However, the available historical datasets span relatively moderate levels of solar activity when compared to previous solar cycles (e.g. Solar Cycles 20-22) and the accuracy of the model may decrease during periods when solar activity exceeds that of the historical datasets. The model performance during future stronger solar cycles can be improved by recomputing the model coefficients using the more active half of the historical data.

The EUVS Model uncertainties for the long-term component are comparable to those from the most recently published FISM update by \cite{thiemann2017maven}.   A key difference between the FISM and EUVS models is that FISM uses a single measurement as an input for each model interval and decomposes the daily average measurement and spectral irradiance predictions into solar cycle and solar rotation components, whereas the EUVS Model uses multiple measurements as inputs for each model interval and does not decompose them according to longer-term solar variability.  This indicates that, at least at 5 nm sampling, daily average irradiance can be accurately predicted without decomposing solar variability into solar cycle and solar rotation components if the model inputs span a broad range of temperatures in the Sun's atmosphere using model coefficients derived from multiple linear regression analysis.  This has important implications for real-time solar spectral irradiance estimation, which does not have the future knowledge required when decomposing solar irradiance variability into solar cycle and solar rotation components.

On the other hand, the EUVS Model uncertainties for the short-term component are a significant improvement over those from FISM.  For example, \cite{thiemann2017maven} reported that the FISM-M flare (i.e. short-term) model uncertainty is 35\% and 70\% in the 13.5 nm and 30.5 nm (1-nm wide) intervals, respectively.  These values should be compared to the 10-15 nm and 30-35 nm interval uncertainties of  26.8\% and 12.7\%, respectively, reported in Table \ref{tab_model_short}.  Although some of the difference in uncertainties is due to the EUVS Model having larger intervals than FISM, by comparing the model measurement comparisons in Figure \ref{fig_flare_scatter} with those for FISM-M in Figure 4 of \cite{thiemann2017maven}, it is clear that a significant part of the uncertainty differences are due to FISM using soft X-ray irradiance variability, which varies by factors of 10-100, to estimate EUV irradiance variability, which varies by tens of percent.  This is most pronounced in the intervals that contain the He II 30.4 nm  emission line.  The uncertainties of the EUVS Model and FISM for the intervals containing the Fe XXIII 13.3 nm  emission line are similar, and likely due to Fe XXIII tending to evolve similarly to the 0.1-0.8 nm band \citep{thiemann2017time}.  In order to reduce model uncertainty, future empirical flare irradiance models should use EUV irradiance measurements as model inputs when possible.  However, it is important to note that \cite{thiemann2018center} showed that the GOES XRS peak emission measure is linearly proportional to peak EUV irradiance for hot-forming emission lines ($T > 9$ MK) during flares, unlike the XRS irradiances, which have a highly non-linear relationship with EUV emissions as discussed above and reported in \cite{thiemann2017maven}.  As such, the XRS emission measure is another alternative for reducing flare irradiance model uncertainty.

The EUVS Model can be improved in a number of ways:  An intermediate non-flaring time-scale could be added to accurately capture variability that occurs at time-scales of a few hours that is not the result of flares as seen, for example, in Figure \ref{fig_flare_example_2012}.   This could be a third term in Equation \ref{eqn_model} at an intermediate time scale, say 2 hours.  The current EUVS Model does not explicitly attempt to model the various EUV emission flare phases.  Figures \ref{fig_flare_example_2012} and  \ref{fig_flare_example_2011} suggest that the $k_i$ coefficients are weighted to capture some of the flare phases accurately.  For example,  model intervals that tend to show an impulsive flare phase are predominantly driven by the impulsive He II 30.4 nm model input, and those intervals that do not tend to be impulsive tend to be predominantly driven by the Fe XXIV 25.3 nm  (nominally the He II EXIS 25.6 nm  line measurement).  Further, from Figure \ref{fig_flare_example_2012}, the dimming phase appears to be accurately modeled for the two flares shown although it is important to note that a false dimming phase tends to appear in the 10-15 nm interval for long duration flares.  The gradual phase delay is not properly modeled using the current EUVS Model algorithm.  \cite{thiemann2017time} showed that by applying a simple differential equation to emission line measurements from hotter EUV lines, the delay and broadening apparent in cooler EUV lines can be predicted if the time difference between the line peaks can be determined independently.  Finally, this study did not consider the degree to which the EUVS Model can reproduce the EUV Late Phase.  The  Fe XV 28.4 nm EUVS measurement does tend to show the EUV Late Phase and, as such, could be used to drive model intervals that also show the EUV Late Phase.

The same methods used to derive the coefficients and uncertainties for the relatively coarse wavelength intervals presented here can be applied to other more common wavelength schemes used in upper atmospheric research such as those introduced in \cite{richards1994euvac} and  \cite{solomon2005solar}.  The broadband intervals of \cite{richards1994euvac} correspond directly with the broadband intervals presented here, and the remaining work involves finding coefficients for the 17 emission line intervals.  The \cite{solomon2005solar} intervals do not match those presented here and also include wavelengths below 5 nm.  As such, producing coefficients for the \cite{solomon2005solar} intervals would require re-sampling the source datasets and expanding the source data to include wavelengths below 5 nm.

The results presented in this paper are for the EXIS instrument on the GOES-16 satellite.  The reader should follow the link in the Acknowledgements section for possible updates, as well as values for other flight models as they become available.    The EXIS dependent model uncertainties, $\sigma_{CC,i}$, are expected to be similar between EXIS models on GOES- 17-19.  As such, it is expected the EUVS Model uncertainties presented here for GOES-16/EXIS will hold for the entire GOES-R series EXIS instruments.  The EUVS Model coefficients are expected to be similar between GOES- 16-19 because the EXIS derived inputs, $P_i(t)$ and $Q_i(t)$, are independent of absolute calibration.  However, $P_i(t)$ and $Q_i(t)$ for GOES- 17-19 should be fit to those from GOES-16 to confirm this, and $k_i$ and $j_i$ should be scaled accordingly if the corresponding slopes differ from unity.

\begin{acknowledgements}
    This work was funded by the NOAA GOES Program contract to the University of Colorado, NNG07HW00C.  The data used in this study are publicly available for download through the following locations:  SDO/EVE data are available at http://lasp.colorado.edu/home/eve/data/data-access/. TIMED/SEE data are available at http://lasp.colorado.edu/home/see/data/. SORCE/SOLSTICE data are available at http://lasp.colorado.edu/home/sorce/data/. Mg II c/w data are available at http://www.iup.uni-bremen.de/gome/gomemgii.html.
    
    Once the EXIS data have been approved for public release, EUVS Model spectra will be publicly available through the National Centers for Environmental Information website, https://www.ngdc.noaa.gov/stp/spaceweather.html.  Additionally, future updates to the model coefficients and uncertainties for GOES-R, as well as future releases of those parameters for GOES-S through GOES-U will be available through https://data.ngdc.noaa.gov/platforms/solar-space-observing-satellites/goes/goesXX/l1b/exis-l1b-sfeu/model$\_$coefficients/ , where XX=16-19.

    \end{acknowledgements}


\bibliography{bib_file_w_doi}

\begin{thebibliography}{37}
\providecommand{\natexlab}[1]{#1}
\providecommand{\url}[1]{\texttt{#1}}
\providecommand{\urlprefix}{URL }
\providecommand{\eprint}[2][]{\url{#2}}

\bibitem[{Amblard et~al.(2008)Amblard, Moussaoui, De~Wit, Aboudarham,
  Kretzschmar, Lilensten, and Auch{\`e}re}]{amblard2008euv}
Amblard, P.-O., S.~Moussaoui, T.~D. De~Wit, J.~Aboudarham, M.~Kretzschmar,
  J.~Lilensten, and F.~Auch{\`e}re, 2008.
\newblock The EUV Sun as the superposition of elementary Suns.
\newblock \emph{Astronomy \& Astrophysics}, \textbf{487}(2), L13--L16.
\newblock \url{https://doi.org/10.1051/0004-6361:200809588}.

\bibitem[{Antia et~al.(2003)Antia, Bhatnagar, and
  Ulmschneider}]{antia2003lectures}
Antia, H., A.~Bhatnagar, and P.~Ulmschneider, 2003.
\newblock Lectures on solar physics, vol. 619.
\newblock Springer Science \& Business Media.
\newblock 10.1007/3-540-36963-5.

\bibitem[{Cessateur et~al.(2011)Cessateur, de~Wit, Kretzschmar, Lilensten,
  Hochedez, and Snow}]{cessateur2011monitoring}
Cessateur, G., T.~D. de~Wit, M.~Kretzschmar, J.~Lilensten, J.-F. Hochedez, and
  M.~Snow, 2011.
\newblock Monitoring the solar UV irradiance spectrum from the observation of a
  few passbands.
\newblock \emph{Astronomy \& Astrophysics}, \textbf{528}, A68.
\newblock \url{http://dx.doi.org/10.1051/0004-6361/201015903}.

\bibitem[{Chamberlin et~al.(2018)Chamberlin, Woods, Didkovsky, Eparvier, Jones
  et~al.}]{chamberlin2018solar}
Chamberlin, P., T.~Woods, L.~Didkovsky, F.~Eparvier, A.~Jones, et~al., 2018.
\newblock Solar Ultraviolet Irradiance Observations of the Solar Flares During
  the Intense September 2017 Storm Period.
\newblock \emph{Space Weather}.
\newblock \url{https://doi.org/10.1029/2018SW001866}.

\bibitem[{Chamberlin et~al.(2007)Chamberlin, Woods, and
  Eparvier}]{Chamberlin2007flare}
Chamberlin, P.~C., T.~N. Woods, and F.~G. Eparvier, 2007.
\newblock Flare irradiance spectral model (FISM): Daily component algorithms
  and results.
\newblock \emph{Space Weather}, \textbf{5}(7).
\newblock \url{https://doi.org/10.1029/2007SW000316}.

\bibitem[{Chamberlin et~al.(2008)Chamberlin, Woods, and
  Eparvier}]{Chamberlin2008flare}
Chamberlin, P.~C., T.~N. Woods, and F.~G. Eparvier, 2008.
\newblock Flare irradiance spectral model (FISM): Flare component algorithms
  and results.
\newblock \emph{Space Weather}, \textbf{6}(5).
\newblock \url{https://doi.org/10.1029/2007SW000372}.

\bibitem[{Chamberlin et~al.(2009)Chamberlin, Woods, Eparvier, and
  Jones}]{chamberlin2009next}
Chamberlin, P.~C., T.~N. Woods, F.~G. Eparvier, and A.~R. Jones, 2009.
\newblock Next generation x-ray sensor (XRS) for the NOAA GOES-R satellite
  series.
\newblock In Solar Physics and Space Weather Instrumentation III, vol. 7438,
  743802. International Society for Optics and Photonics.
\newblock \url{https://doi.org/10.1117/12.826807}.

\bibitem[{Davies(1990)}]{davies1990ionospheric}
Davies, K., 1990.
\newblock Ionospheric radio.
\newblock 31. IET.
\newblock 10.1049/PBEW031E.

\bibitem[{Dudok~de Wit et~al.(2009)Dudok~de Wit, Kretzschmar, Lilensten, and
  Woods}]{dudok2009finding}
Dudok~de Wit, T., M.~Kretzschmar, J.~Lilensten, and T.~Woods, 2009.
\newblock Finding the best proxies for the solar UV irradiance.
\newblock \emph{Geophysical Research Letters}, \textbf{36}(10).
\newblock \url{https://doi.org/10.1029/2009GL037825}.

\bibitem[{Dudok~de Wit et~al.(2008)Dudok~de Wit, Kretzschmar, Aboudarham,
  Amblard, Auch{\`e}re, and Lilensten}]{de2008solar}
Dudok~de Wit, T.~D., M.~Kretzschmar, J.~Aboudarham, P.-O. Amblard,
  F.~Auch{\`e}re, and J.~Lilensten, 2008.
\newblock Which solar EUV indices are best for reconstructing the solar EUV
  irradiance?
\newblock \emph{Advances in Space Research}, \textbf{42}(5), 903--911.
\newblock \url{https://doi.org/10.1016/j.asr.2007.04.019}.

\bibitem[{Efron(1979)}]{efron1979bootstrap}
Efron, B., 1979.
\newblock Bootstrap methods: another look at the jackknife annals of statistics
  7: 1--26.
\newblock \emph{Ann. Statist.}, \textbf{7}(1), 1--26.
\newblock 10.1214/aos/1176344552.

\bibitem[{Eparvier et~al.(2009)Eparvier, Crotser, Jones, McClintock, Snow, and
  Woods}]{eparvier2009extreme}
Eparvier, F.~G., D.~Crotser, A.~R. Jones, W.~E. McClintock, M.~Snow, and T.~N.
  Woods, 2009.
\newblock The extreme ultraviolet sensor (EUVS) for GOES-R.
\newblock In Solar Physics and Space Weather Instrumentation III, vol. 7438,
  743804. International Society for Optics and Photonics.
\newblock \url{https://doi.org/10.1117/12.826445}.

\bibitem[{Heath and Schlesinger(1986)}]{heath1986mg}
Heath, D.~F., and B.~M. Schlesinger, 1986.
\newblock The Mg 280-nm doublet as a monitor of changes in solar ultraviolet
  irradiance.
\newblock \emph{Journal of Geophysical Research: Atmospheres}, \textbf{91}(D8),
  8672--8682.
\newblock \url{https://doi.org/10.1029/JD091iD08p08672}.

\bibitem[{Hinteregger(1981)}]{hinteregger1981representations}
Hinteregger, H., 1981.
\newblock Representations of solar EUV fluxes for aeronomical applications.
\newblock \emph{Advances in Space Research}, \textbf{1}(12), 39--52.
\newblock \url{https://doi.org/10.1016/0273-1177(81)90416-6}.

\bibitem[{Jachhia(1959)}]{jacchia1959two}
Jachhia, L., 1959.
\newblock Two Atmospheric Effects in the Orbital Acceleration of Artificial
  Satellites.
\newblock \emph{Nature}, \textbf{183}(4660), 526--527.
\newblock 10.1038/183526a0, \urlprefix\url{https://doi.org/10.1038/183526a0}.

\bibitem[{Kretzschmar et~al.(2006)Kretzschmar, Lilensten, and
  Aboudarham}]{kretzschmar2006retrieving}
Kretzschmar, M., J.~Lilensten, and J.~Aboudarham, 2006.
\newblock Retrieving the solar EUV spectral irradiance from the observation of
  6 lines.
\newblock \emph{Advances in Space Research}, \textbf{37}(2), 341--346.
\newblock \url{https://doi.org/10.1016/j.asr.2005.02.029}.

\bibitem[{Lilensten et~al.(2007)Lilensten, de~Wit, Amblard, Aboudarham,
  Auch{\`e}re, and Kretzschmar}]{lilensten2007recommendation}
Lilensten, J., T.~D. de~Wit, P.-O. Amblard, J.~Aboudarham, F.~Auch{\`e}re, and
  M.~Kretzschmar, 2007.
\newblock Recommendation for a set of solar EUV lines to be monitored for
  aeronomy applications.
\newblock In Annales Geophysicae, vol.~25, 1299--1310.
\newblock \url{https://doi.org/10.5194/angeo-25-1299-2007}.

\bibitem[{McClintock et~al.(2005)McClintock, Rottman, and
  Woods}]{mcclintock2005solar}
McClintock, W.~E., G.~J. Rottman, and T.~N. Woods, 2005.
\newblock Solar--Stellar Irradiance Comparison Experiment II (SOLSTICE II):
  Instrument concept and design.
\newblock \emph{Solar Physics}, \textbf{230}(1-2), 225--258.
\newblock \url{https://doi.org/10.1007/s11207-005-7432-x}.

\bibitem[{Mendillo et~al.(1974)Mendillo, Klobuchar, Fritz, Da~Rosa, Kersley
  et~al.}]{mendillo1974behavior}
Mendillo, M., J.~Klobuchar, R.~Fritz, A.~Da~Rosa, L.~Kersley, et~al., 1974.
\newblock Behavior of the ionospheric F region during the great solar flare of
  August 7, 1972.
\newblock \emph{Journal of Geophysical Research}, \textbf{79}(4), 665--672.
\newblock \url{https://doi.org/10.1029/JA079i004p00665}.

\bibitem[{Qian et~al.(2010)Qian, Burns, Chamberlin, and
  Solomon}]{qian2010flare}
Qian, L., A.~G. Burns, P.~C. Chamberlin, and S.~C. Solomon, 2010.
\newblock Flare location on the solar disk: Modeling the thermosphere and
  ionosphere response.
\newblock \emph{Journal of Geophysical Research: Space Physics},
  \textbf{115}(A9).
\newblock \url{https://doi.org/10.1029/2009JA015225}.

\bibitem[{Richards et~al.(1994)Richards, Fennelly, and
  Torr}]{richards1994euvac}
Richards, P., J.~Fennelly, and D.~Torr, 1994.
\newblock EUVAC: A solar EUV flux model for aeronomic calculations.
\newblock \emph{Journal of Geophysical Research: Space Physics},
  \textbf{99}(A5), 8981--8992.
\newblock \url{https://doi.org/10.1029/94JA00518}.

\bibitem[{Solomon and Qian(2005)}]{solomon2005solar}
Solomon, S.~C., and L.~Qian, 2005.
\newblock Solar extreme-ultraviolet irradiance for general circulation models.
\newblock \emph{Journal of Geophysical Research: Space Physics},
  \textbf{110}(A10).
\newblock \url{https://doi.org/10.1029/2005JA011160}.

\bibitem[{Suess et~al.(2016)Suess, Snow, Viereck, and Machol}]{suess2016solar}
Suess, K., M.~Snow, R.~Viereck, and J.~Machol, 2016.
\newblock Solar Spectral Proxy Irradiance from GOES (SSPRING): a model for
  solar EUV irradiance.
\newblock \emph{Journal of Space Weather and Space Climate}, \textbf{6}, A10.
\newblock \url{https://doi.org/10.1051/swsc/2016003}.

\bibitem[{Taylor(1997)}]{taylor1997introduction}
Taylor, J., 1997.
\newblock Introduction to error analysis, the study of uncertainties in
  physical measurements.
\newblock University Science Books, Sausalito, CA.

\bibitem[{Thiemann et~al.(2018)Thiemann, Chamberlin, Eparvier, and
  Epp}]{thiemann2018center}
Thiemann, E., P.~Chamberlin, F.~Eparvier, and L.~Epp, 2018.
\newblock Center-to-Limb Variability of Hot Coronal EUV Emissions During Solar
  Flares.
\newblock \emph{Solar Physics}, \textbf{293}(2), 19.
\newblock \url{https://doi.org/10.1007/s11207-018-1244-2}.

\bibitem[{Thiemann et~al.(2017{\natexlab{a}})Thiemann, Chamberlin, Eparvier,
  Templeman, Woods, Bougher, and Jakosky}]{thiemann2017maven}
Thiemann, E.~M., P.~C. Chamberlin, F.~G. Eparvier, B.~Templeman, T.~N. Woods,
  S.~W. Bougher, and B.~M. Jakosky, 2017{\natexlab{a}}.
\newblock The MAVEN EUVM model of solar spectral irradiance variability at
  Mars: Algorithms and results.
\newblock \emph{Journal of Geophysical Research: Space Physics},
  \textbf{122}(3), 2748--2767.
\newblock \url{https://doi.org/10.1002/2016JA023512}.

\bibitem[{Thiemann et~al.(2017{\natexlab{b}})Thiemann, Eparvier, and
  Woods}]{thiemann2017time}
Thiemann, E.~M., F.~G. Eparvier, and T.~N. Woods, 2017{\natexlab{b}}.
\newblock A time dependent relation between EUV solar flare light-curves from
  lines with differing formation temperatures.
\newblock \emph{Journal of Space Weather and Space Climate}, \textbf{7}, A36.
\newblock \url{https://doi.org/10.1051/swsc/2017037}.

\bibitem[{Tobiska and Eparvier(1998)}]{tobiska1998euv97}
Tobiska, W.~K., and F.~Eparvier, 1998.
\newblock EUV97: Improvements to EUV irradiance modeling in the soft X-rays and
  FUV.
\newblock \emph{Solar Physics}, \textbf{177}(1-2), 147--159.
\newblock \url{https://doi.org/10.1023/A:10049314}.

\bibitem[{Tobiska et~al.(2000)Tobiska, Woods, Eparvier, Viereck, Floyd, Bouwer,
  Rottman, and White}]{tobiska2000solar2000}
Tobiska, W.~K., T.~Woods, F.~Eparvier, R.~Viereck, L.~Floyd, D.~Bouwer,
  G.~Rottman, and O.~White, 2000.
\newblock The SOLAR2000 empirical solar irradiance model and forecast tool.
\newblock \emph{Journal of Atmospheric and Solar-Terrestrial Physics},
  \textbf{62}(14), 1233--1250.
\newblock \url{https://doi.org/10.1016/S1364-6826(00)00070-5}.

\bibitem[{Torr and Torr(1985)}]{torr1985ionization}
Torr, M.~R., and D.~Torr, 1985.
\newblock Ionization frequencies for solar cycle 21: Revised.
\newblock \emph{Journal of Geophysical Research: Space Physics},
  \textbf{90}(A7), 6675--6678.
\newblock \url{https://doi.org/10.1029/JA090iA07p06675}.

\bibitem[{Van~Huffel(1989)}]{van1989extended}
Van~Huffel, S., 1989.
\newblock The extended classical total least squares algorithm.
\newblock \emph{Journal of computational and applied mathematics},
  \textbf{25}(1), 111--119.
\newblock \url{https://doi.org/10.1016/0377-0427(89)90080-0}.

\bibitem[{Veronig et~al.(2002)Veronig, Temmer, Hanslmeier, Otruba, and
  Messerotti}]{veronig2002temporal}
Veronig, A., M.~Temmer, A.~Hanslmeier, W.~Otruba, and M.~Messerotti, 2002.
\newblock Temporal aspects and frequency distributions of solar soft X-ray
  flares.
\newblock \emph{Astronomy \& Astrophysics}, \textbf{382}(3), 1070--1080.
\newblock \url{https://doi.org/10.1051/0004-6361:20011694}.

\bibitem[{Viereck et~al.(2007)Viereck, Hanser, Wise, Guha, Jones, McMullin,
  Plunket, Strickland, and Evans}]{viereck2007solar}
Viereck, R., F.~Hanser, J.~Wise, S.~Guha, A.~Jones, D.~McMullin, S.~Plunket,
  D.~Strickland, and S.~Evans, 2007.
\newblock Solar extreme ultraviolet irradiance observations from GOES: design
  characteristics and initial performance.
\newblock In Solar Physics and Space Weather Instrumentation II, vol. 6689,
  66890K. International Society for Optics and Photonics.
\newblock \url{https://doi.org/10.1117/12.734886}.

\bibitem[{Woods et~al.(2010)Woods, Eparvier, Hock, Jones, Woodraska
  et~al.}]{woods2010extreme}
Woods, T., F.~Eparvier, R.~Hock, A.~Jones, D.~Woodraska, et~al., 2010.
\newblock Extreme Ultraviolet Variability Experiment (EVE) on the Solar
  Dynamics Observatory (SDO): Overview of science objectives, instrument
  design, data products, and model developments.
\newblock In B.~T. P.~Chamberlin, W.~Pesnell, ed., The Solar Dynamics
  Observatory, 115--143. Springer.
\newblock \url{https://doi.org/10.1007/s11207-011-9841-3}.

\bibitem[{Woods et~al.(2005)Woods, Eparvier, Bailey, Chamberlin, Lean, Rottman,
  Solomon, Tobiska, and Woodraska}]{woods2005solar}
Woods, T.~N., F.~G. Eparvier, S.~M. Bailey, P.~C. Chamberlin, J.~Lean, G.~J.
  Rottman, S.~C. Solomon, W.~K. Tobiska, and D.~L. Woodraska, 2005.
\newblock Solar EUV Experiment (SEE): Mission overview and first results.
\newblock \emph{Journal of Geophysical Research: Space Physics},
  \textbf{110}(A1).
\newblock \url{https://doi.org/10.1029/2004JA010765}.

\bibitem[{Woods et~al.(2011)Woods, Hock, Eparvier, Jones, Chamberlin
  et~al.}]{woods2011new}
Woods, T.~N., R.~Hock, F.~Eparvier, A.~R. Jones, P.~C. Chamberlin, et~al.,
  2011.
\newblock New solar extreme-ultraviolet irradiance observations during flares.
\newblock \emph{The Astrophysical Journal}, \textbf{739}(2), 59.
\newblock \url{https://doi.org/10.1088/0004-637X/739/2/59}.

\bibitem[{Woods et~al.(2000)Woods, Tobiska, Rottman, and
  Worden}]{woods2000improved}
Woods, T.~N., W.~K. Tobiska, G.~J. Rottman, and J.~R. Worden, 2000.
\newblock Improved solar Lyman $\alpha$ irradiance modeling from 1947 through
  1999 based on UARS observations.
\newblock \emph{Journal of Geophysical Research: Space Physics},
  \textbf{105}(A12), 27,195--27,215.
\newblock \url{https://doi.org/10.1029/2000JA000051}.

\end{thebibliography}

  \newpage
  
  \appendix

\section{EUVS Model Coefficients} \label{app_l1b}

 \begin{table}[ht]
  \scriptsize
 \caption{EUVS Measurement Reference Values.}  \label{tab_ch_min}
 \centering
 \begin{tabular}{c c c c c c c c} 
 \hline
  $X_{25.6}$ & $X_{28.4}$ & $X_{30.4}$ & $X_{117.5}$ & $X_{121.6}$ & $X_{133.5}$ & $X_{140.5}$& $X_{280}$\\
 \hline
 2.23e-05 & 2.713e-05 & 3.82e-04 & 8.245e-05 & 5.95e-03 & 1.72e-04 & 1.15e-04 & 0.305\\
\hline
 \end{tabular}
 \end{table}

 \begin{table}
 \scriptsize
 \caption{Long-Term Model Coefficients.}  \label{tab_model_long_ap}
 \centering
 \begin{tabular}{l c c c c c c c c c} 
 \hline
 Bin (nm) & $E_{\lambda,0}$ & $j_{25.6}$ & $j_{28.4}$ & $j_{30.4}$ & $j_{117.5}$ & $j_{121.6}$ & $j_{133.5}$ & $j_{140.5}$& $j_{280}$\\
 \hline
 5-10 & 1.86e-05 & 6.91e-06 & 1.83e-06 & 0.00e+00 & 1.29e-05 & 0.00e+00 & 0.00e+00 & 1.45e-05 & 0.00e+00\\
10-15 & 9.33e-06 & 2.44e-06 & 0.00e+00 & 0.00e+00 & 2.12e-06 & 0.00e+00 & 0.00e+00 & 0.00e+00 & 0.00e+00\\
15-20 & 5.42e-05 & 1.28e-05 & 1.77e-07 & 0.00e+00 & 0.00e+00 & 1.18e-06 & 0.00e+00 & 0.00e+00 & 0.00e+00\\
20-25 & 2.51e-05 & 1.03e-05 & 2.06e-06 & 0.00e+00 & 0.00e+00 & 1.37e-05 & 0.00e+00 & 0.00e+00 & 0.00e+00\\
25-30 & 2.10e-05 & 5.53e-06 & 6.32e-06 & 0.00e+00 & 0.00e+00 & 0.00e+00 & 0.00e+00 & 0.00e+00 & 0.00e+00\\
30-35 & 1.12e-04 & 1.47e-06 & 2.73e-06 & 7.21e-05 & 7.37e-06 & 0.00e+00 & 0.00e+00 & 0.00e+00 & 0.00e+00\\
35-40 & 2.94e-05 & 3.55e-07 & 4.53e-06 & 6.95e-06 & 1.97e-07 & 0.00e+00 & 0.00e+00 & 0.00e+00 & 3.71e-07\\
40-45 & 6.93e-06 & 2.55e-07 & 2.51e-07 & 2.29e-06 & 0.00e+00 & 0.00e+00 & 0.00e+00 & 0.00e+00 & 4.77e-06\\
45-50 & 1.15e-05 & 1.96e-07 & 1.77e-07 & 5.69e-06 & 8.95e-07 & 1.10e-06 & 0.00e+00 & 0.00e+00 & 7.74e-06\\
50-55 & 7.74e-06 & 3.86e-07 & 1.82e-07 & 4.83e-06 & 2.38e-07 & 1.82e-06 & 0.00e+00 & 0.00e+00 & 5.69e-06\\
55-60 & 1.75e-05 & 2.43e-07 & 0.00e+00 & 2.86e-06 & 4.44e-06 & 9.39e-07 & 0.00e+00 & 0.00e+00 & 1.66e-05\\
60-65 & 1.91e-05 & 4.46e-07 & 0.00e+00 & 5.78e-06 & 2.32e-07 & 0.00e+00 & 0.00e+00 & 2.63e-06 & 1.51e-05\\
65-70 & 5.51e-06 & 3.18e-08 & 0.00e+00 & 1.39e-06 & 8.22e-07 & 0.00e+00 & 0.00e+00 & 0.00e+00 & 3.78e-06\\
70-75 & 7.15e-06 & 2.69e-08 & 0.00e+00 & 3.25e-07 & 1.88e-06 & 0.00e+00 & 0.00e+00 & 0.00e+00 & 5.40e-06\\
75-80 & 1.56e-05 & 0.00e+00 & 0.00e+00 & 0.00e+00 & 0.00e+00 & 0.00e+00 & 0.00e+00 & 0.00e+00 & 1.50e-05\\
80-85 & 1.87e-05 & 0.00e+00 & 0.00e+00 & 0.00e+00 & 0.00e+00 & 0.00e+00 & 0.00e+00 & 0.00e+00 & 3.18e-05\\
85-90 & 3.30e-05 & 2.36e-07 & 0.00e+00 & 8.80e-06 & 0.00e+00 & 5.91e-06 & 0.00e+00 & 0.00e+00 & 4.21e-05\\
90-95 & 2.95e-05 & 2.09e-07 & 0.00e+00 & 7.91e-06 & 7.31e-07 & 0.00e+00 & 6.60e-06 & 0.00e+00 & 3.08e-05\\
95-100 & 3.49e-05 & 5.52e-07 & 0.00e+00 & 3.10e-06 & 2.37e-06 & 5.01e-06 & 0.00e+00 & 4.16e-06 & 3.47e-05\\
100-105 & 4.45e-05 & 5.43e-07 & 8.62e-08 & 1.43e-05 & 2.08e-06 & 0.00e+00 & 2.80e-06 & 7.43e-06 & 4.50e-05\\
105-110 & 1.67e-05 & 1.08e-07 & 0.00e+00 & 5.26e-06 & 5.59e-06 & 0.00e+00 & 0.00e+00 & 0.00e+00 & 1.85e-05\\
110-115 & 1.83e-05 & 4.57e-08 & 0.00e+00 & 1.95e-06 & 1.54e-06 & 0.00e+00 & 0.00e+00 & 0.00e+00 & 2.67e-05\\
117-127 & 6.72e-04 & 0.00e+00 & 0.00e+00 & 0.00e+00 & 0.00e+00 & 6.57e-04 & 0.00e+00 & 0.00e+00 & 0.00e+00\\
\hline
 \end{tabular}
 \end{table}

 \begin{table}
 \scriptsize
 \caption{Short-Term Model Coefficients.}  \label{tab_model_short_ap}
 \centering
 \begin{tabular}{l c c c c c c c c} 
 \hline
 Bin (nm) & $k_{25.6}$ & $k_{28.4}$ & $k_{30.4}$ & $k_{117.5}$ & $k_{121.6}$ & $k_{133.5}$ & $k_{140.5}$& $k_{280}$\\
5-10 & 3.66e-05 & 2.12e-05 & 0.00e+00 & 0.00e+00 & 0.00e+00 & 0.00e+00 & 0.00e+00 & 0.00e+00\\
10-15 & 1.12e-04 & 3.97e-05 & 0.00e+00 & 0.00e+00 & 0.00e+00 & 0.00e+00 & 0.00e+00 & 0.00e+00\\
15-20 & 4.36e-05 & 0.00e+00 & 7.74e-06 & 0.00e+00 & 0.00e+00 & 0.00e+00 & 0.00e+00 & 0.00e+00\\
20-25 & 1.27e-05 & 6.16e-06 & 1.12e-05 & 0.00e+00 & 4.76e-06 & 0.00e+00 & 0.00e+00 & 0.00e+00\\
25-30 & 2.07e-05 & 1.37e-05 & 3.92e-06 & 0.00e+00 & 0.00e+00 & 0.00e+00 & 0.00e+00 & 0.00e+00\\
30-35 & 0.00e+00 & 0.00e+00 & 1.11e-04 & 0.00e+00 & 2.89e-05 & 0.00e+00 & 0.00e+00 & 0.00e+00\\
35-40 & 0.00e+00 & 1.06e-05 & 2.73e-05 & 0.00e+00 & 2.47e-05 & 0.00e+00 & 0.00e+00 & 0.00e+00\\
40-45 & 9.23e-07 & 2.93e-06 & 4.41e-06 & 0.00e+00 & 1.47e-06 & 0.00e+00 & 0.00e+00 & 0.00e+00\\
45-50 & 0.00e+00 & 3.29e-06 & 1.29e-05 & 0.00e+00 & 0.00e+00 & 0.00e+00 & 0.00e+00 & 0.00e+00\\
50-55 & 1.16e-06 & 1.71e-06 & 9.20e-06 & 0.00e+00 & 0.00e+00 & 0.00e+00 & 0.00e+00 & 0.00e+00\\
55-60 & 0.00e+00 & 0.00e+00 & 3.10e-05 & 0.00e+00 & 0.00e+00 & 0.00e+00 & 0.00e+00 & 0.00e+00\\
60-65 & 0.00e+00 & 0.00e+00 & 2.03e-05 & 0.00e+00 & 0.00e+00 & 0.00e+00 & 0.00e+00 & 0.00e+00\\
65-70 & 0.00e+00 & 0.00e+00 & 0.00e+00 & 0.00e+00 & 1.07e-05 & 0.00e+00 & 0.00e+00 & 0.00e+00\\
70-75 & 0.00e+00 & 0.00e+00 & 2.88e-06 & 0.00e+00 & 1.25e-05 & 0.00e+00 & 0.00e+00 & 0.00e+00\\
75-80 & 0.00e+00 & 0.00e+00 & 2.02e-05 & 0.00e+00 & 2.25e-05 & 0.00e+00 & 0.00e+00 & 0.00e+00\\
80-85 & 0.00e+00 & 2.15e-05 & 1.22e-05 & 0.00e+00 & 3.54e-05 & 0.00e+00 & 0.00e+00 & 0.00e+00\\
85-90 & 1.83e-05 & 7.77e-06 & 7.58e-06 & 0.00e+00 & 5.01e-05 & 0.00e+00 & 0.00e+00 & 0.00e+00\\
90-95 & 0.00e+00 & 0.00e+00 & 3.23e-05 & 0.00e+00 & 1.79e-05 & 0.00e+00 & 0.00e+00 & 0.00e+00\\
95-100 & 0.00e+00 & 0.00e+00 & 1.32e-04 & 0.00e+00 & 1.02e-04 & 0.00e+00 & 0.00e+00 & 0.00e+00\\
100-105 & 0.00e+00 & 0.00e+00 & 1.02e-04 & 0.00e+00 & 0.00e+00 & 0.00e+00 & 0.00e+00 & 0.00e+00\\
105-110 & 0.00e+00 & 0.00e+00 & 8.16e-06 & 0.00e+00 & 1.00e-05 & 1.57e-05 & 2.87e-06 & 0.00e+00\\
110-115 & 0.00e+00 & 0.00e+00 & 8.78e-06 & 0.00e+00 & 0.00e+00 & 9.33e-06 & 5.51e-06 & 0.00e+00\\
117-127 & 0.00e+00 & 0.00e+00 & 0.00e+00 & 0.00e+00 & 7.54e-04 & 0.00e+00 & 0.00e+00 & 0.00e+00\\
 \hline
  \hline
  \end{tabular}
 \end{table}


\end{document}